\documentclass[preprint,amsmath,amssymb,pre]{revtex4-1}
\usepackage{graphicx}
\usepackage{graphics}
\usepackage{dcolumn}
\usepackage{bm}
\usepackage{lineno}
\usepackage{hyperref}
\usepackage{bm}
\usepackage{graphicx}
\usepackage{graphics}
\usepackage{amsmath}
\usepackage{cases}
\usepackage{array}
\usepackage{amssymb}
\usepackage{float}
\usepackage{booktabs}
\usepackage{multirow}
\usepackage{siunitx}
\usepackage{subfig}
\usepackage{tabularx}
\usepackage{amsmath}
\usepackage{cleveref}
\usepackage{color}
\usepackage[utf8]{inputenc}
\usepackage{natbib}

\graphicspath{{./FIG_PY/}}

\begin{document}
\preprint{APS/123-QED}

\title{ A fractional step lattice Boltzmann model for two phase flows with large density differences}

\author{Chunhua Zhang}
\author{Zhaoli Guo}
\email{zlguo@hust.edu.cn}
\affiliation{
State Key Laboratory of Coal Combustion, Huazhong University of Science and Technology, Wuhan 430074, China}
\author{Yibao Li}
\affiliation{
School of Mathematics and Statistics, Xi'an Jiaotong University, Xi'an 710049, China}
\date{\today}

\begin{abstract}
In this paper, \textcolor{blue}{a fractional step lattice Boltzmann method is proposed to model two-phase flows with large density differences by solving Cahn-Hilliard phase-field equation and the incompressible Navier-Stokes equations.}
In order to  maintain a hyperbolic tangent property of the interface profile and conserve the volume, an interfacial profile correction term and a flux correction term are added into the original Cahn-Hilliard equation  respectively. By using a fractional step scheme, the modified Cahn-Hilliard equation is split into two sub-equations. One is solved in the framework of lattice Boltzmann equation method. The other is  solved by the finite difference method. Compared with the previous lattice Boltzmann methods, the proposed method is able to maintain the order parameter within  a physically meaningful range, which is conductive to track the interface accurately.  In addition, the multi-relaxation-time collision model and a high-order compact selective filter operation are employed to enhance the numerical stability. \textcolor{blue}{The proposed method can simulate two-phase fluid flows with the density ratio up to $1000$.}
In order to validate the accuracy and capability of the method,  several benchmark problems, including single vortex deform of a circle, translation of a drop, Laplace-Young law, capillary wave and rising bubble with large density ratios, are presented. \textcolor{blue}{The results are in good agreement with the analytical solutions and the data in the literature for the investigated benchmarks.
}
 \vspace{10pt}
\end{abstract}
\maketitle

\section{Introduction}
Over the years, the phase-field model has emerged as a reliable and versatile method for multiphase flows~\cite{anderson1998diffuse,ding2007diffuse,lee2010lattice,yue2004diffuse,jacqmin1999calculation,li2016multi}.
In this framework, the bulk fluid is represented as an order parameter with constant values. The interface between two phase fluids is described as a finite transition region where phase variable can continuously vary between their values in the bulk fluids. This leads to the two main advantages of diffuse-interface model. One advantage is natural capability of capturing complex topological changes. Another feature is its rigorous thermodynamic basis, such as, most models satisfy a nonlinear stability relationship by using a non-convex free-energy function. The evolution of the order variable is governed by the phase-field equations, including  Cahn-Hilliard (CH) and Allen Cahn (AC) equations. To model two-phase flows, the phase field equation should couple with a fluid flow governed by  the modified Navier-Stokes (NS) equations with a surface force, leading to the so-called Navier-Stokes/Cahn-Hilliard (NSCH) or Navier-Stokes/Allen-Cahn (NSAC) system.
Based on the phase-field theory for multiphase flows, a great number of numerical methods have been developed~\cite{jacqmin1999calculation,yue2004diffuse,fakhari2010phase,
zu2013phase,shen2015decoupled}. Among these methods, the phase-field-based lattice Boltzmann equation (LBE) method has received extensive attention because of its some features~\cite{chen1998lattice,lee2005stable,fakhari2010phase,zu2013phase,liang2014phase}.

The LBE method is based on simplified mesoscopic kinetic equation with discrete-velocity distribution function instead of discretizations of the macroscopic governing equations. However, the macroscopic governing equations can be recovered from the lattice Boltmzann equation through the Champan-Enskog expansion only if the equilibrium function and the source term are designed properly. \textcolor{blue}{At present, most of the LBE methods for two phase flows are developed based on the NSCH system.} To correctly recover the desired NSCH equations, great efforts have been made~\cite{zheng2005lattice,lee2005stable,zu2013phase,liang2014phase,inamuro2016improved}.
For example, Zheng \emph{et al.}\cite{zheng2005lattice} added a spatial difference term of the distribution function as the source term to recovery the CH equation correctly. Liang \emph{et al.}\cite{liang2014phase} proposed a time-derivative term related to the order parameter and the velocity in the source term to ensure the correct CH equation.  Li \emph{et al.}~\cite{li2012additional} presented an additional interfacial force to recover the target momentum equation correctly by considering the continuity equation being nonzero near the interface. \textcolor{blue}{
However, it is still not satisfactory for the capability of capturing the interfaces and modeling two-phase flows with large density contrasts.
\textcolor{blue}{In particular, it is a challenging task for the NSCH system because the CH equation is a fourth-order partial differential equation and the system of NSCH equations is very stiff.}
To model large-density-ratio flows, several approaches have been developed.}
For example, Wang \emph{et al.}~\cite{wang2015lattice} proposed a LBE flux solver for two-phase flows with large density ratios, in which the CH equation is solved by a stable high-order WENO difference scheme. Lee \emph{et al.}~\cite{lee2005stable} proposed a stable mixing difference scheme for the force terms in their model to achieve large density ratios.  Inamuro \emph{et al}~\cite{inamuro2004lattice} proposed a free-energy method for large density ratio flow, in which an additional pressure poisson equation is solved iteratively to enforce the divergence-free condition. Kim \emph{et al.}~\cite{kim2015lattice} employed a filtering operation for the pressure field to enhance the numerical stability in addition to improving the divergence-free velocity by reducing the incompressibility error.
\textcolor{blue}{Although these methods can simulate the large-density-ratio flows, another important problem has not received much attention.} The value of the order parameter defined as the difference between the volume fractions of two phases should theoretically be $[-1,1]$.
 However, the numerical solutions of the CH equation have maxima and minima values that are larger than $\phi=1$ and smaller than $\phi=-1$, respectively, due to flow and diffusion.  These extreme values can give rise to nonphysical fluid characteristics. In particular, for multiphase flows with large density ratios, this phenomenon is more pronounced leading to numerical instability. In order to address this problem, a cutoff technique for the order parameter is widely used once the order parameter exceeds the theoretical maximum and minimum values~\cite{inamuro2004lattice,ren2016improved,shen2015decoupled}. However, this method violates the mass conservation. Recently, Li~\emph{et al.}~\cite{li2016phase} proposed a modified CH equation with an interfacial profile correction term based on the equilibrium profile obtained in the thermodynamically derived phase-field model. The added term is used to enforce the phase-field profile to be a hyperbolic tangent profile.
As a result, the values of order parameter in the bulk of each component are almost $\pm 1$.
Based on the work of Li~\emph{et al.}, Zhang and Ye~\emph{et al}~\cite{zhang2017flux} further proposed a flux-corrected term by considering removing the bulk diffusion motion which may contribute to the loss of volume. And the  modified CH equation performed a good performance in terms of eliminating the shrinkage and coarsening effects.
\textcolor{blue}{Although the hydrodynamic equation is not considered in their work, this modified CH equation is also potential for studying two-phase fluid flows with large density ratios.}

In addition, the Allen Cahn phase field equation is also proposed for capturing the interface of two phases~\cite{chiu2011conservative,sun2007sharp,geier2015conservative}. \textcolor{blue}{Compared with the CH equation with a fourth-order derivative, the AC equation is easier to be solved because only second-order partial differential equation are involved.}
By recovering the AC and NS equations though the Champman-Enskog analysis, several LBE methods of multiphase flows with large density ratios have been proposed~\cite{liang2018phase,fakhari2017diffuse,joshi2018positivity}.

In this work, our objective is to develop a LBE method to model two-phase flows with large density ratios based on the NSCH system.
\textcolor{blue}{To do this,
the modified CH equation mentioned above with a convection term is employed instead of the original convection CH equation. The resulting equation is solved by a fractional step scheme efficiently. To improve the numerical stability, the multi-relaxation-time (MRT) collision model is employed and the source term distribution function is redesigned to recover the governing equations correctly. In addition, a filtering operation is applied to the pressure field and velocity field as well.}

The rest of the present paper is organized as follows. In section 2, we describe the modified phase-field equation and Navier-Stokes equations of two incompressible fluids with different densities. In section 3,  the MRT-LBE model is constructed to recover the macroscopic governing equations correctly. The proposed model will be verified in section 4. Finally, a brief summary is given in Section 5.

\section{Governing equations}
\subsection{The modified Cahn-Hilliard equation}
In the diffuse interface model, the evolution of an order parameter $\phi$ is generally governed by a convective CH equation,
\begin{equation}\label{eq:originalCH}
  \partial_t \phi+\nabla\cdot(\phi\bm u)=M\nabla^2\mu_{\phi},
\end{equation}
where $\bm u$ is the velocity, $M$ is the mobility and $\mu_{\phi}$ is the chemical potential defined below.
 The CH equation is derived by minimizing the free energy of the system, which can be described by the Helmholtz free energy functional~\cite{abels2012thermodynamically},
\begin{equation}\label{eq:free_energy}
\Psi(\phi)=\beta (\phi^2-1)^2+\frac{\kappa}{2}\left|\nabla \phi\right|^2,
\end{equation}
where $\beta$ and $\kappa$ are constant that depend on the surface tension $\sigma$ and the interface thickness W, i.e., $\kappa=\frac{3}{8}\sigma W$ and $\beta=\frac{3\sigma}{4W}$.
A variational procedure applied to the total free energy will yield the chemical potential~\cite{jacqmin2000contact,yan2007lattice},
\begin{equation}\label{eq:chemical}
\mu_{\phi} \equiv \frac{\delta \Psi}{\delta \phi}=4\beta(\phi^3-\phi)-\kappa\nabla^2\phi.
\end{equation}
Consider a one-dimensional interface at the equilibrium condition,  one can obtain the equilibrium profile for the order parameter, 
\begin{equation}\label{eq:equilibrium_profile}
\phi^{eq}(\bm r)=\tanh\left(\frac{2\bm r}{W}\right),
\end{equation}
 where $\bm r$ is the coordinate normal to the interface. Based on Eq.~(\ref{eq:equilibrium_profile}), the interface profile between the two phases should be a hyperbolic tangent profile at equilibrium.
\textcolor{blue}{Motivated by this idea, Li~\emph{et al.}~\cite{li2016phase} proposed an interface correction term to maintain the hyperbolic tangent profile of the interface~\cite{li2016phase}. More recently,
a flux correction term is developed to maintain conserve the volume~\cite{zhang2017flux}. By considering these two correction terms, the original convection CH equation can be modified to the following form,}
\begin{equation}\label{eq:final_mCHE}
\partial_t \phi+\nabla\cdot (\phi\bm u)=M\nabla^2\mu_{\phi}+\nabla\cdot \bm J_{\phi}+ \nabla\cdot\bm q,
\end{equation}
 where $\bm J_{\phi}=\left(|\nabla\phi|-\frac{2(1-\phi^2)}{W}\right)\bm n $ is used to enforce the interface profile to be $\tanh\left(\frac{2r}{W}\right)$, $\bm q=(-M\nabla\mu_{\phi}\cdot\bm n)\bm n$ is used to remove the bulk diffusion, and $\bm n=\frac{\nabla\phi}{|\nabla\phi|}$ is an outward pointing normal vector.
 Instead of directly solving Eq~(\ref{eq:final_mCHE}) at each time step, a fractional step scheme is employed, which consists of the following two sequential steps:
\begin{align}\label{eq:step1}
\partial_t \phi+\nabla\cdot (\phi\bm u)&= M\nabla^2\mu_{\phi},\\
\label{eq:step2}
\partial_{t} \phi&=\nabla\cdot \bm J_d.
\end{align}
where $\bm J_d= \bm J_{\phi}+\bm q$. Eq~(\ref{eq:step1}) is solved by the LBE method and Eq~(\ref{eq:step2}) can be efficiently solved by  the finite difference method, which will be presented below.

\subsection{The incompressible Navier-Stokes equations}
We here consider two-phase flows of immiscible incompressible fluids with different densities.  The governing equations in the conservation form are as follows~\cite{abels2017diffuse,abels2012thermodynamically},
\begin{subequations}
\begin{align}
\label{eq:divgenceVelocity}
\nabla \cdot \bm{u}&=0, \\
\label{eq:momentum}
\partial_t (\rho\bm u) + \nabla\cdot(\rho \bm u \bm u) &= -\nabla p_h
+\nabla\cdot (2\mu D(\bm u))+\nabla \cdot(\bm J\bm u)+\bm F_s+\bm F_g,
\end{align}
\end{subequations}
with
\begin{equation}\label{eq:rho}
\rho =\rho_1\frac{1+\phi}{2}+\rho_2\frac{1-\phi}{2},
\end{equation}
where $\mu$ is the dynamic viscosity, $D(\bm u)=\frac{1}{2}(\nabla\bm u+\nabla \bm u^T)$ denotes the rate of deformation tensor, $\bm I$ is Kronecker delta, $\rho_1$ and $\rho_2$ are the component densities,
$p_h$ is the pressure, $\bm J=\frac{\partial\rho}{\partial\phi}M\nabla\mu_{\phi}$ is the flux due to the diffusion, $\bm F_s=-\kappa\nabla\cdot(\nabla \phi\otimes\nabla\phi)$ describes the capillary force exerted to the fluids by the interface, $\bm F_g=\rho \bm g$ is the gravity force with $\bm g$ representing the gravitational acceleration.
The above system is thermodynamically consistent  and satisfies an energy dissipation law. \textcolor{blue}{In addition to the surface force above, several forms of the surface force have been used in the literature~\cite{kim2005continuous}. In this study, we employ the following formulation~\cite{starovoitov1994model},
\begin{equation}\label{eq:surfaceforce}
\bm F_s=\kappa\nabla\cdot(|\nabla \phi|^2-\nabla \phi\otimes\nabla\phi).
\end{equation}
Then, the pressure is redefined as $p'=p_h+\kappa |\nabla \phi|^2$.}

The momentum equaiton can also be expressed as the following  equivalent nonconservative form
\begin{equation}\label{eq:reformNS}
\partial_t \bm u+\nabla \cdot (\bm u\bm u) = -\nabla p
         +\nabla\cdot 2\nu D(\bm u)
         +\frac{\bm F_p+\bm F_s+\bm F_{\nu}+\bm F_{m}+ \bm F_g}{\rho},
\end{equation}
where $\nu=\mu/\rho$ is the kinematic viscosity, $\bm F_p=-p\nabla \rho$, $\bm F_{\nu}=\nu (\nabla\bm u+ \nabla \bm u^{T})\cdot\nabla\rho$
, $\bm F_m=\bm J\cdot \nabla \bm u$ and $p=p'/\rho$.

\section{Phase-field-Based Lattice Boltzmann method for incompressible two-phase flows }
\subsection{The MRT-LBE model for the Cahn-Hilliard equation}
 The standard form of discrete Boltzmann equation with MRT collision model for the CH equation can be written as
\begin{equation}\label{eq:distribution_CH}
f_i( \bm{x}+\bm{c_i} \delta t,t+\delta t)- f_i(\bm{x},t)=
 -\Lambda_{ij}^{f}(f_j(\bm{x},t)-f_j^{eq}(\bm{x},t))+ \left(\bm I-\frac{\Lambda_{ij}^f}{2}\right) F_j\delta t,
\end{equation}
where $f_i(\bm{x},t)$ is the distribution functions for the order parameter field with a discrete velocity $\bm c_i$ at point $\bm x$ and time $t$, $f_i^{eq}(\bm x,t)$ is the corresponding equilibrium state, $\delta t$ is the time step, $\Lambda_{ij}^f$ is the collision operate defined below, $\bm I$ is the unit matrix and ${F}_j$ is the source term.
The equilibrium distribution functions $f_i^{eq}(\bm{x},t)$ is defined as~\cite{zhang2018high}
\begin{equation}\label{eq:feq_CH}
  f_i^{eq}=\left\{
  \begin{aligned}
  &\phi+(\omega_0-1)\eta\mu- \omega_0 \phi \frac{\bm u^2}{2c_s^2},& & \text {$i=0$}\\
  &\omega_i\eta\mu+\omega_i \phi \left(\frac{\bm c_i \cdot \bm u}{c_s^2}+\frac{(\bm c_i\cdot \bm u)^2}{2c_s^4}-\frac{\bm u^2}{2c_s^2}\right),&  & \text {$i \neq 0$}
  \end{aligned}
  \right.
\end{equation}
where $\omega_i$ is the weighting coefficient, $c_s$ is the sound speed and $\eta$ is an adjustable parameter for a given mobility. Because of only two-dimensional simulations being considered, the two-dimensional nine-velocity (D2Q9) structures is used to solve flow fields. The discrete particle velocity $\bm c_i$ in D2Q9 can be defined as
\begin{equation}\label{eq:D2Q9}
\bm c_i=
\left\{
  \begin{aligned}
&(0,0)e,& & \text{i=0} \\
&(\cos[(i-1)\pi/2],\sin[(i-1)\pi/2])e,& & \text{i=1,\ldots,4} \\
&\sqrt{2}(\cos[(i-5)\pi/2+\pi/4],\sin[(i-5)\pi/2+\pi/4])e,&  & \text{i=5,\ldots,8},
\end{aligned}
\right.
\end{equation}
where $ e= \delta x / \delta t=\delta y/\delta t$ with $\delta x$ and $\delta y$ representing the lattice spacing in the x-direction and y-direction respectively.
 In this study, both $\delta x$ ($\delta y$) and $\delta t$ are fixed at unity.
The corresponding weighting coefficient in each direction is given by $\omega_0=4/9$, $\omega_{1,\ldots,4}=1/9$, $\omega_{1,\ldots, 8}=1/36$ and $c_s=1/\sqrt{3}$.

In the MRT-LBE method, the collision with the forcing term effect is executed in momentum space by multiplying a transformation matrix $\bm{\text{M}}$ while the propagation of the MRT model remains in velocity space. Thus, the collision operate can be written as,
\begin{equation}\label{eq:collison}
\Lambda_{ij}^f=\bm{\text{M}}^{-1}\bm{\text{S}}^f\bm{\text{M}},
\end{equation}
with
\begin{equation}\label{eq:D2Q9MRT}
\bm{\text{M}}=\left[ \begin{array}{ccccccccc}
  1  & 1  & 1  & 1 & 1 & 1 & 1  &1  &1\\
  -4 & -1 & -1 & -1 & -1 & 2 & 2  &2  &2\\
  4 & -2  & -2 & -2 & -2 & 1 & 1  &1  &1\\
  0 & 1   & 0  & -1 & 0 & 1 & -1  &-1  &1\\
  0 & -2  & 0  & 2 & 0 & 1 & -1  &-1  &1\\
  0 & 0 & 1  & 0 & -1 & 1 & 1  &-1  &-1\\
  0 & 0 & -2 & 0 & 2 & 1 & 1  &-1  &-1\\
  0 & 1 & -1 & 1 & -1 & 0 & 0  &0  &0\\
  0 & 0 & 0  & 0 & 0 & 1 & -1  &1  &-1
\end{array}
\right],
\end{equation}
Here $\bm{\text{S}}^f$ is a diagonal relaxation matrix,
\begin{equation}\label{eq:CH_Sf}
\bm{\text{S}}^{f}=\text{diag}(s_0^f,s_1^f,s_2^f,s_3^f,s_4^f,s_5^f,s_6^f,s_7^f,s_8^f),
\end{equation}
where each element represents the inverse of a relaxation time for the
distribution function $f_i$ as it is relaxed to the equilibrium distribution function in moment space. For the stability, $0\le s_i^f\le 2$ must be satisfied.
The parameters $s_3^f$ and $s_5^f$ are related to the relaxation time $\tau_f$ in the single-relaxation-time (SRT) model, i.e., $s_3^f=s_5^f=1/\tau_f$. If all elements equal to each other, the MRT model will reduce to the SRT model. The source term $F_i$ in Eq.(\ref{eq:distribution_CH}) is given by~\cite{zhang2018high}
\begin{equation}\label{eq:Fi}
F_i=\omega_i  \frac{\phi\bm c_i\cdot\left(\partial_{t}\bm u+\bm u\cdot\nabla \bm u\right)}{c_s^2}.
\end{equation}
Though the Chapman-Enskog analysis, the mobility is defined as
\begin{equation}\label{eq:mobility}
m=\eta c_s^2(\tau_f-0.5)\delta t.
\end{equation}
The order parameter can be calculated by
\begin{equation}\label{eq:compute_phi}
\bar{\phi}=\sum_i f_i.
\end{equation}
In general, the values of  $\bar{\phi}$ not strictly attain in $[-1,1]$. To maintain the physical meaningful values of the order parameter, Eq.~(\ref{eq:step2}) is solved by the finite difference method thereafter. To be specific, a third-order total variation diminishing Runge-Kutta scheme is used for the temporal discretization,
\begin{align}\label{eq:TVD-3order}
\phi^{(1)}&=\bar{\phi}+\delta t L(\bar{\phi}),\\
\phi^{(2)}&=\frac{3}{4}\bar{\phi}+ \frac{1}{4}\left(\phi^{(1)}+\delta t L(\phi^{(1)})\right),\\
\phi &=\frac{1}{3}\bar{\phi}+\frac{2}{3}\left(\phi^{(2)}+\delta t L(\phi^{(2)})\right),
\end{align}
where $L(\phi)=\nabla \cdot \bm J_d$, $\phi^{(1)}$ and $\phi^{(2)}$ are the intermediate values of the order parameter. Here, the four-order center difference scheme is applied to discretize the right hand terms,
\begin{equation}\label{eq:fourOrder_FDM}
\begin{aligned}
\frac{\partial \phi^{*}}{\partial  x}&=\frac{\phi^{*}(x-2\delta x,y)-8\phi^{*}(x-\delta x,y)+8\phi^{*}(x+\delta x,y)-\phi^{*}(i+2\delta x,\delta y)}{12\delta x},\\
\frac{\partial \phi^{*}}{\partial  y}&=\frac{\phi^{*}(x,y-2\delta y)-8\phi^{*}(x,y-\delta y)+8\phi^{*}(x,j+\delta y)-\phi^{*}(x,y-2\delta y)}{12\delta y},
\end{aligned}
\end{equation}
where $\phi^{*}$ represents $\bar{\phi}, \phi^{(1)}$ and $\phi^{(2)}$.

\subsection{The MRT-LBE model for the incompressible Navier-Stokes equations}
The discrete Boltzmann equation with MRT collision model for hydrodynamic equations can be written  as~\cite{he1999lattice}
\begin{equation}\label{eq:distribution_NS}
g_i( \bm{x}+\bm{c_i} \delta t,t+\delta t)- g_i(\bm{x},t)=
 -\Lambda_{ij}^g(g_j(\bm{x},t)-g_j^{eq}(\bm{x},t))+\left(\bm I-\frac{\Lambda_{ij}^g}{2}\right) G_j\delta_t ,
\end{equation}
where $g_i(\bm x,t)$ is a pressure distribution function with a discrete velocity $\bm c_i$ at point $\bm x$ and time $t$, $\Lambda_{ij}^g$ is the collision operator for the pressure distribution function, $g_i^{eq}(\bm x,t)$ is the corresponding equilibrium distribution function and $G_j$ is the source term.  Similar to Eq.~(\ref{eq:collison}), $\Lambda_{ij}^g$ can be expressed as $\bm{\text{M}}^{-1}\bm{\text{S}}^{g}\bm {\text{M}}$, where $\bm{\text{S}}^{g}=\text{diag}(s_0^g, s_1^g, s_2^g,s_3^g,s_4^g,s_5^g,s_6^g,s_7^g,s_8^g )$. Here $s_7^g$ and $s_8^g$ is related to the relaxation time $\tau_g$ in the SRT version, i.e., $s_7^g=s_8^g=1/\tau_g$.
The equilibrium distribution function can be defined as
\begin{equation}\label{eq:geq}
g_i^{eq}=\omega_i \left(\frac{p}{c_s^2}+\frac{\bm c_i\cdot\bm u}{c_s^2}+\frac{(\bm c_i\cdot \bm u)^2}{2c_s^4}-\frac{\bm u^2}{2c_s^2}\right).
\end{equation}
Note that the first moment of the equilibrium distribution functions above is
based on the fluid velocity $\bm u$, not the momentum $ \rho \bm u$.
Compared with the momentum-based formulation, the velocity-based formulation can remove differentiation-by-part error for the divergence-free condition~\cite{kim2015lattice,zu2013phase}.
To recover the momentum equation exactly though the Champan-Enskog expansion, the source term $G_i$ in Eq.~(\ref{eq:distribution_NS}) is defined as~\cite{guo2002discrete},
\begin{equation}\label{eq:Gi}
G_i=\omega_i\frac{1}{\rho}\left(
\frac{\bm c_i \cdot\bm F}{c_s^2}+
\frac{\bm u(\bm F-\nabla p) :(\bm c_i\bm c_i -c_s^2\bm I)}{c_s^4}
 \right),
\end{equation}
where $\bm F=\bm F_p+\bm F_s+\bm F_{\nu}+\bm F_{m}+\bm F_g$.
 Taking the zero- and first order moments of the distribution function $g_i$, the macroscopic quantities $\bm u$ and $p$ are calculated by
\begin{equation}\label{Eq:p}
p= c_s^2\sum_{i} g_i,
\end{equation}
\begin{equation}\label{Eq:u}
\bm{u}=\sum_i g_i \bm c_i+\frac{\bm F}{2\rho}\delta t.
\end{equation}
Note that the deviatoric stress tensor in $\bm F_{\nu}$ can be obtained in terms of the distribution function in addition to using the  finite difference calculation. For the MRT model,  $\nabla\bm u+\nabla\bm u^T$ can be calculated by
\begin{equation}\label{eq:Fmu}
\nabla\bm u+\nabla\bm u^T=\frac{(s_7^g-s_1^g)(\bm F\cdot \bm u)\bm I-s_7^g(\bm F\bm u+\bm u\bm F)}{2\rho c_s^2}-
\frac{\sum_i\bm c_i\bm c_i(\bm M^{-1}\bm S^g\bm M)_{ij}(f_j-f_j^{eq})}{c_s^2\delta t}.
\end{equation}
However, it is clear that obtaining the velocity in Eq.~(\ref{Eq:u}) requires implicit calculation. Instead of implicit calculation, this equation is solved iteratively until $\sum |\bm u^{n+1}-\bm u^n|/\sum |\bm u^{n}|<\zeta$ is satisfied in the whole domain, where $\zeta$ is a given error threshold.
And we found that only few steps are required to maintain numerical stability.

The  viscosity $\mu$ is given by
\begin{equation}\label{eq:dynamic_vis}
\mu =\mu_1\frac{1+\phi}{2}+\mu_2\frac{1-\phi}{2},
\end{equation}
where $\mu_1$ and $\mu_2$ are the viscosities of the two fluids, respectively. Once the viscosity of the fluid is obtained, the relaxation time is computed by
$\tau_g=\mu/(\rho c_s^2\delta t)+0.5$.
Through the Chapman-Enskog analysis, the governing equation in Eq.~(\ref{eq:reformNS}) can be recovered correctly.~(see Appendix.~\ref{eq:derivationNS} for details.)

The force terms in Eq.~(\ref{Eq:u}) involve the first-order derivatives and the Laplacian, which need to be carefully discretized. Like most used, the following isotropic discretization is adopted,
\begin{equation}\label{eq:dicretized}
\begin{aligned}
  \nabla\Phi(\bm x,t)&=\sum_{i\neq0}\frac{\omega_{i}\bm c_{i} \Phi(\bm x+\bm c_i\delta t,t)}{c_s^2\delta t},\\
  \nabla^2\Phi(\bm x,t)&=\sum_{i\neq0}\frac{2\omega_{i} (\Phi(\bm x+\bm c_i\delta t,t)-\Phi(\bm x,t))}{c_s^2\delta t^2},\\
  \end{aligned}
\end{equation}
where $\Phi$ is an arbitrary function.
In addition, the finite-difference schemes may generate grid-to-grid oscillations leading to numerical instability. A spatial selective filter is therefore used to eliminate grid-to-grid oscillations~\cite{tam1993dispersion,ricot2009lattice}. After each time step, the variable $\bm \chi=(p,\bm u)$ is filtered by the following high-order selective filter~\cite{chao2011filter,berland2007high},
\begin{equation}\label{eq:filter}
\bm \chi (\bm x)=\bm \chi(\bm x)-a\sum_{\alpha=1}^{D}\sum_{j=-P}^{Q} d_m \bm \chi(\bm x_{\alpha}+j\delta x),
\end{equation}
where $P$ and $Q$ are the number of points of the stencil, $D$ is the space dimension, $a$ is a constant between 0 and 1, which describes the intensity of the filter and $d_m=d_{-m}$ is a filter coefficient. The filter coefficients are given in Appendix~\ref{selectiveFilter}.

\section{Numerical Results and discussion}
In this section, we will validate the proposed MRT-LBE method through several numerical benchmarks, including  single vortex deformation of a circle, translation of a drop,
 Laplace-Young law, capillary wave and a rising bubble with large density ratio. The results of each test will compared with the analytical solutions and the data in existing literature.

\subsection{Single vortex deformation of a circular droplet}
\begin{figure}[H]
\centering
\subfloat[]{\includegraphics[width=0.5\textwidth]{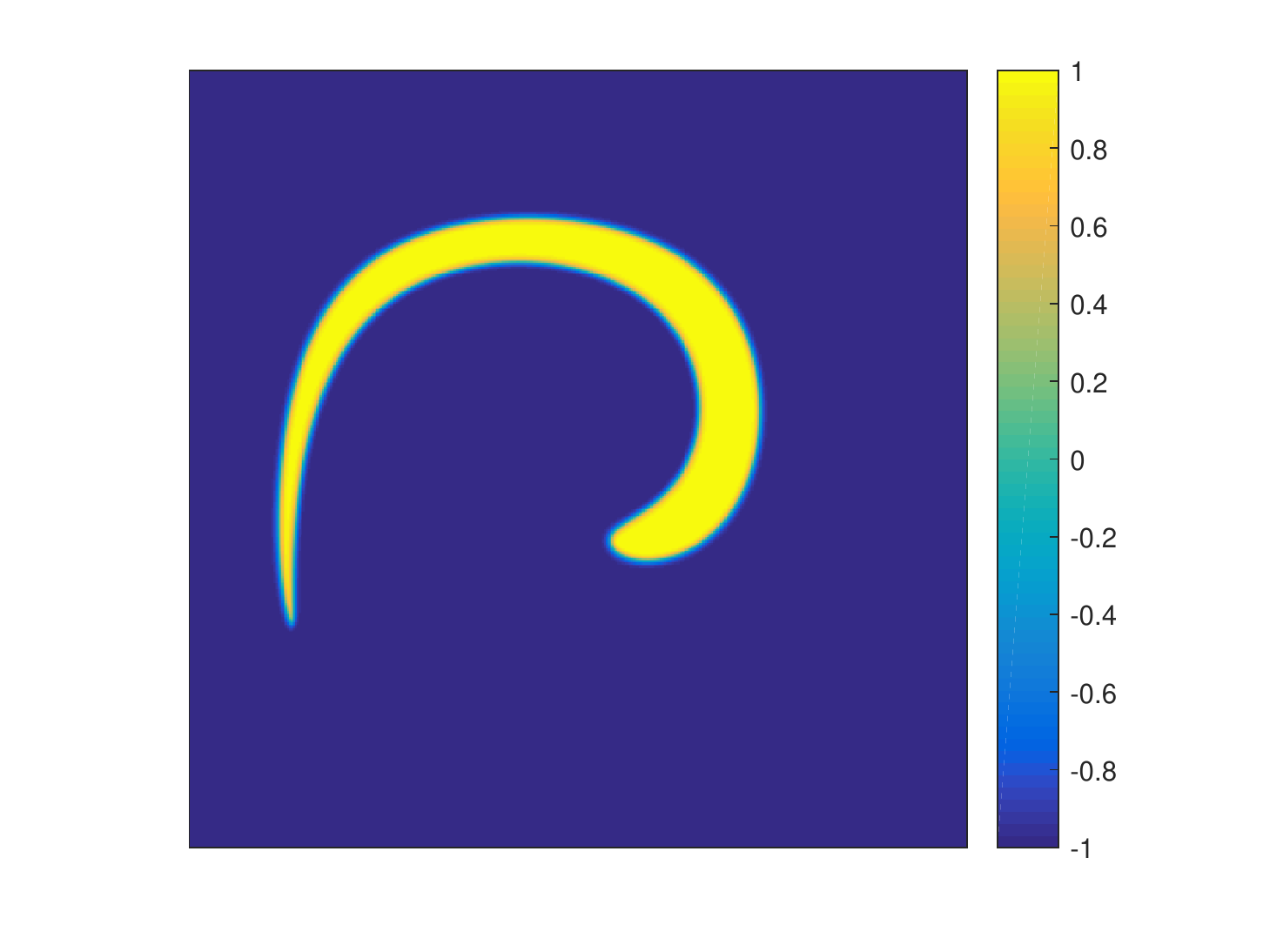}}~
\subfloat[]{\includegraphics[width=0.5\textwidth]{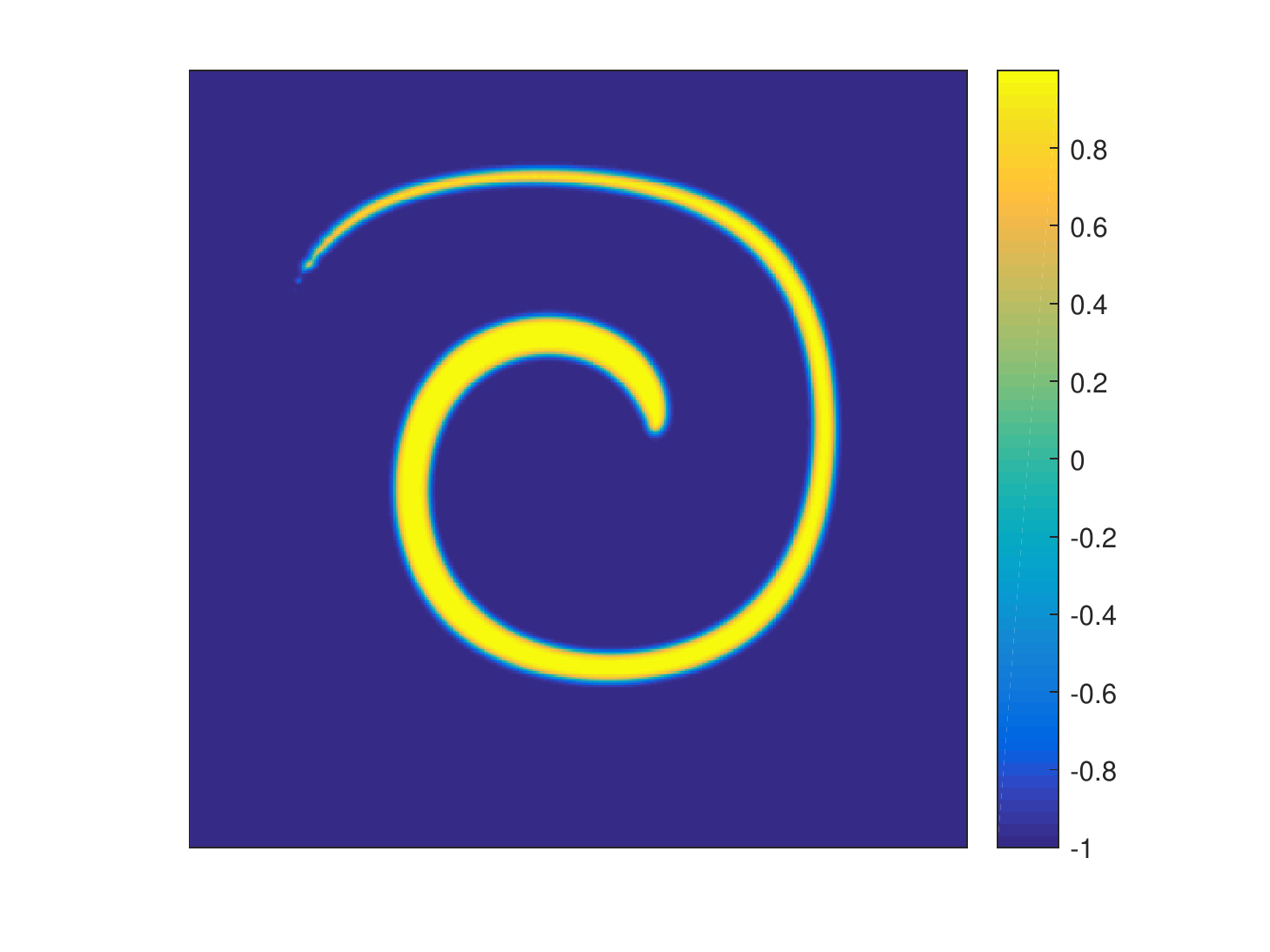}}\\
\subfloat[]{\includegraphics[width=0.5\textwidth]{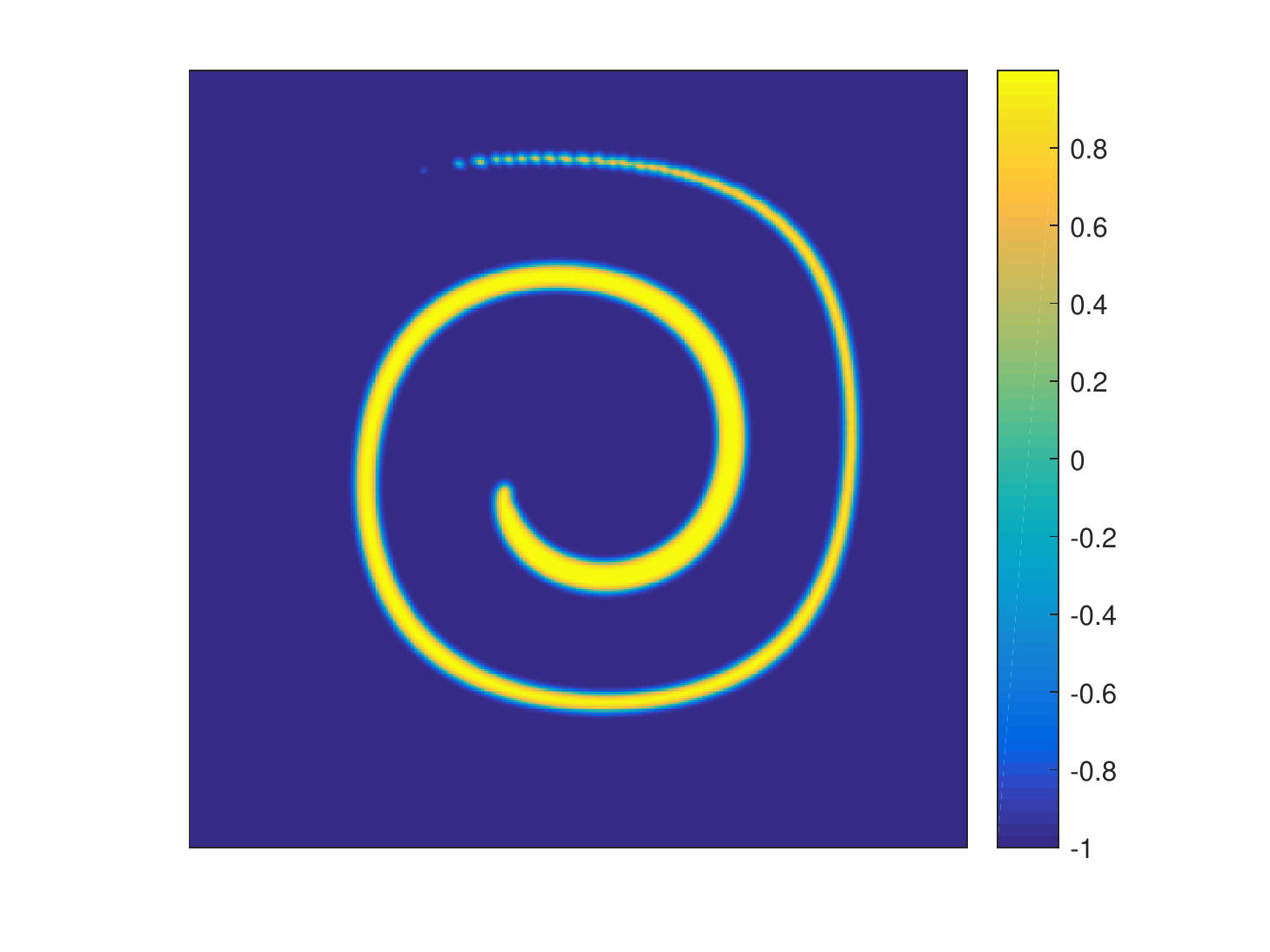}}~
\subfloat[]{\includegraphics[width=0.5\textwidth]{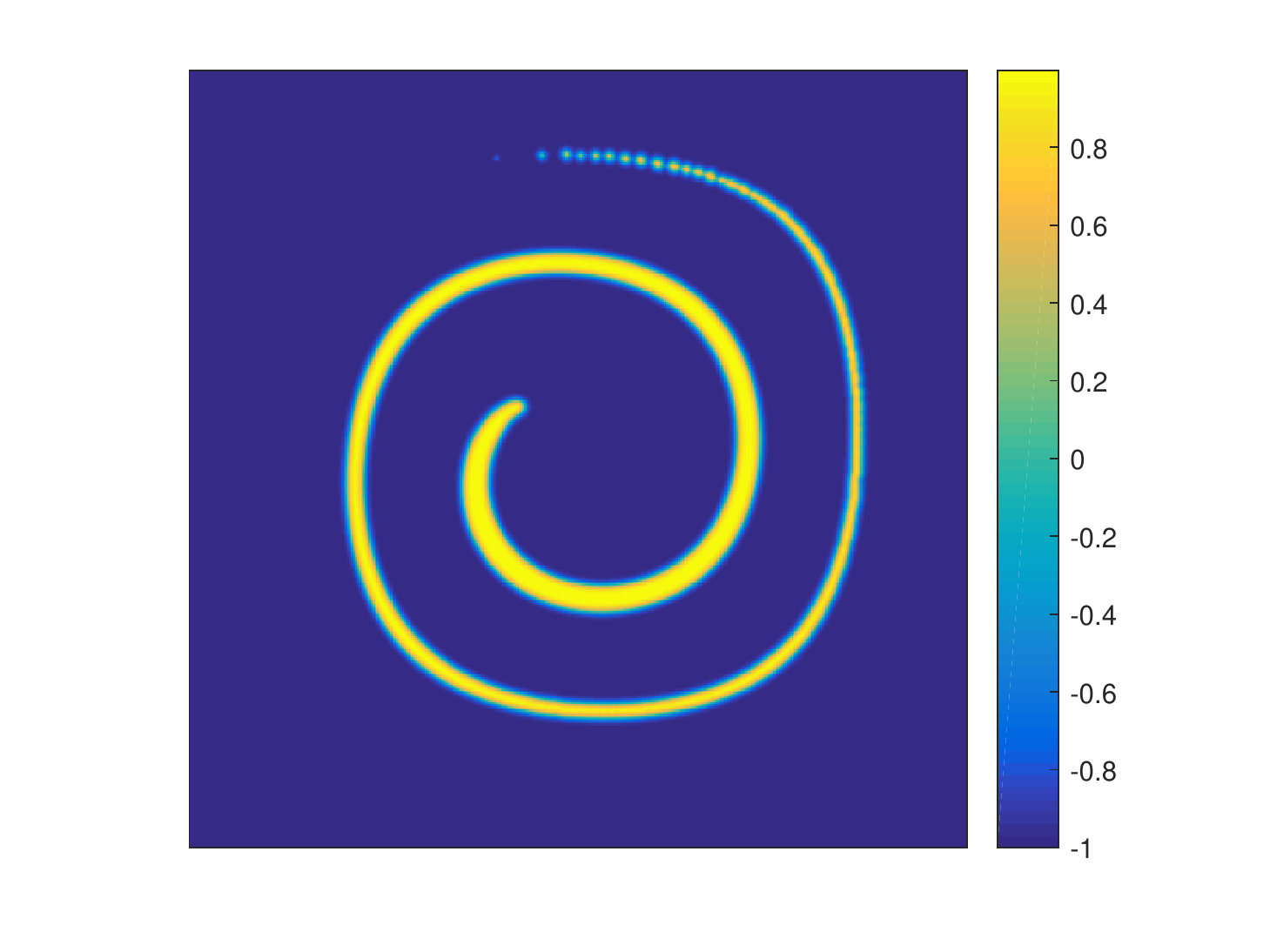}}
\caption{Single vortex deformation of a circular droplet obtained by the present method at n=4. (a) $t=0.2T$, (b) $t=0.4T$, (c) $t=0.8T$, (d) $t=T$.}.
\label{fig:vortex}
\end{figure}

\begin{figure}[H]
\centering
\subfloat{\includegraphics[width=0.5\textwidth]{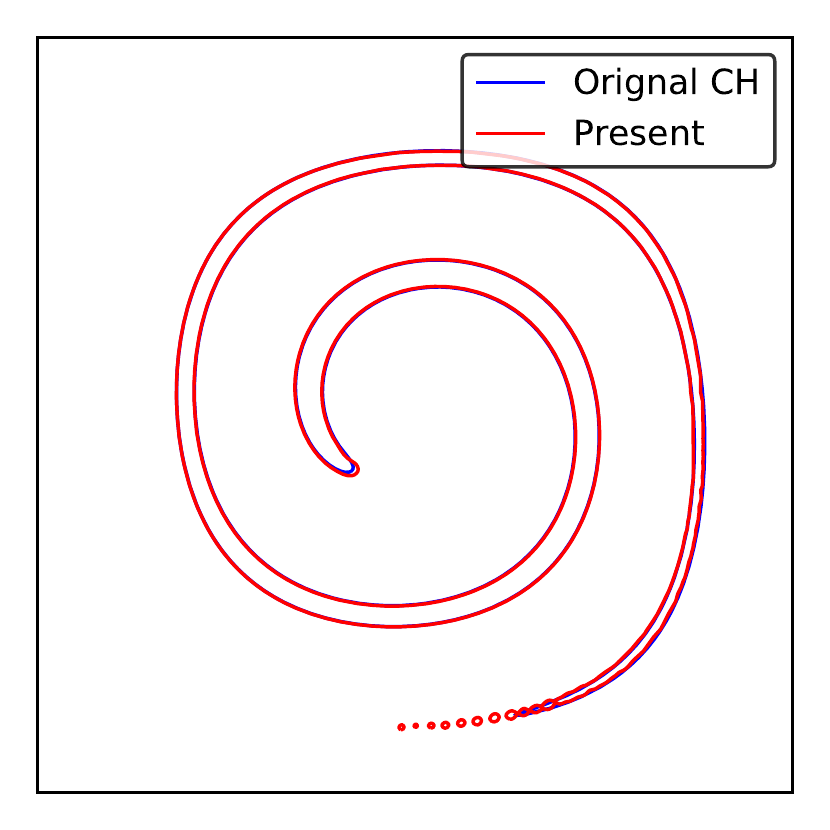}}
\caption{Comparison between the interface shapes obtained by the original CH method (blue line) and those obtained by the present method (red line) at $t=T, n=4$.}
\label{fig:vortex_compareShapeHalf}
\end{figure}

\begin{figure}[H]
\centering
\subfloat[]{\includegraphics[width=0.5\textwidth]{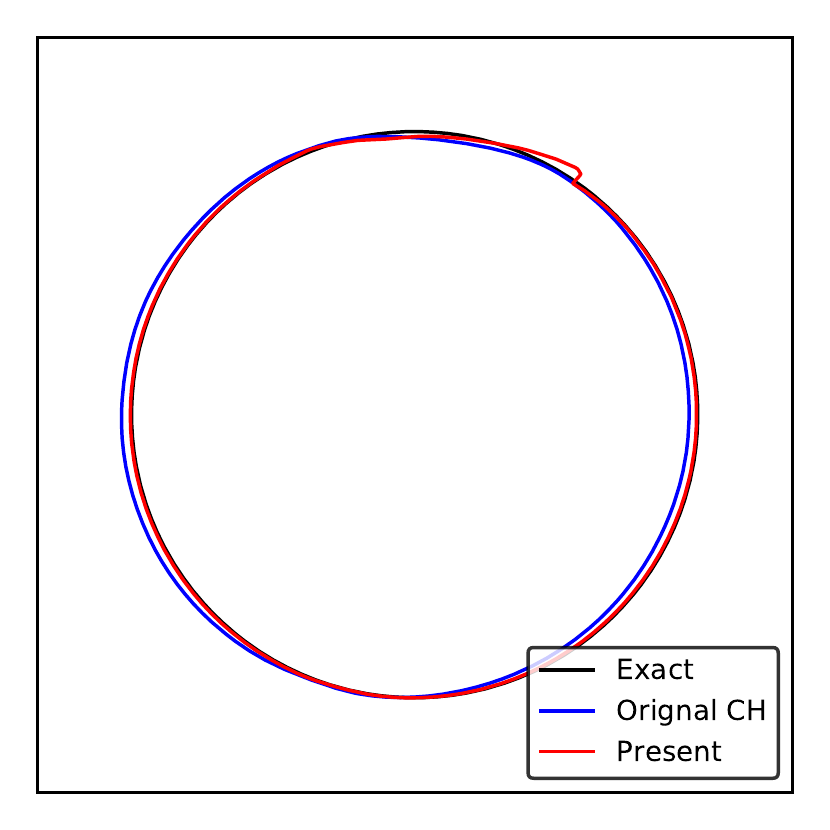}}~
\subfloat[]{\includegraphics[width=0.5\textwidth]{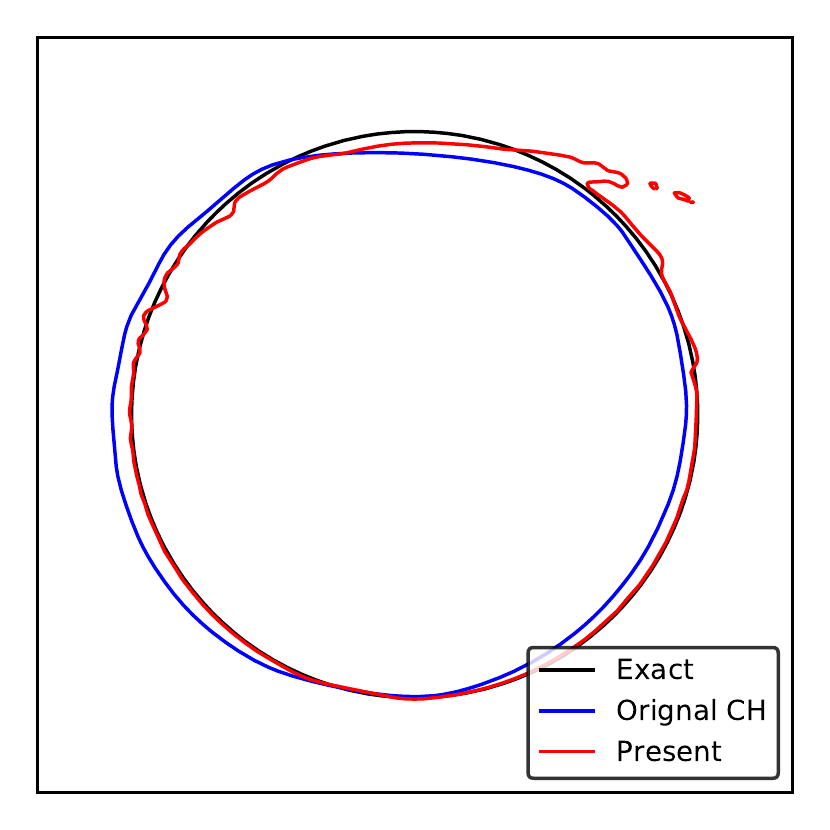}}

\caption{Comparison between the interface shapes obtained by the original CH method (blue line) and those obtained by the present method (red line) at $t=2T$. (a) $n=2$ and (b) $n=4$. The black line denotes the initial interface shapes.}
\label{fig:vortex_compareShapeFinal}
\end{figure}

\begin{figure}[H]
\centering
\subfloat{\includegraphics[width=0.5\textwidth]{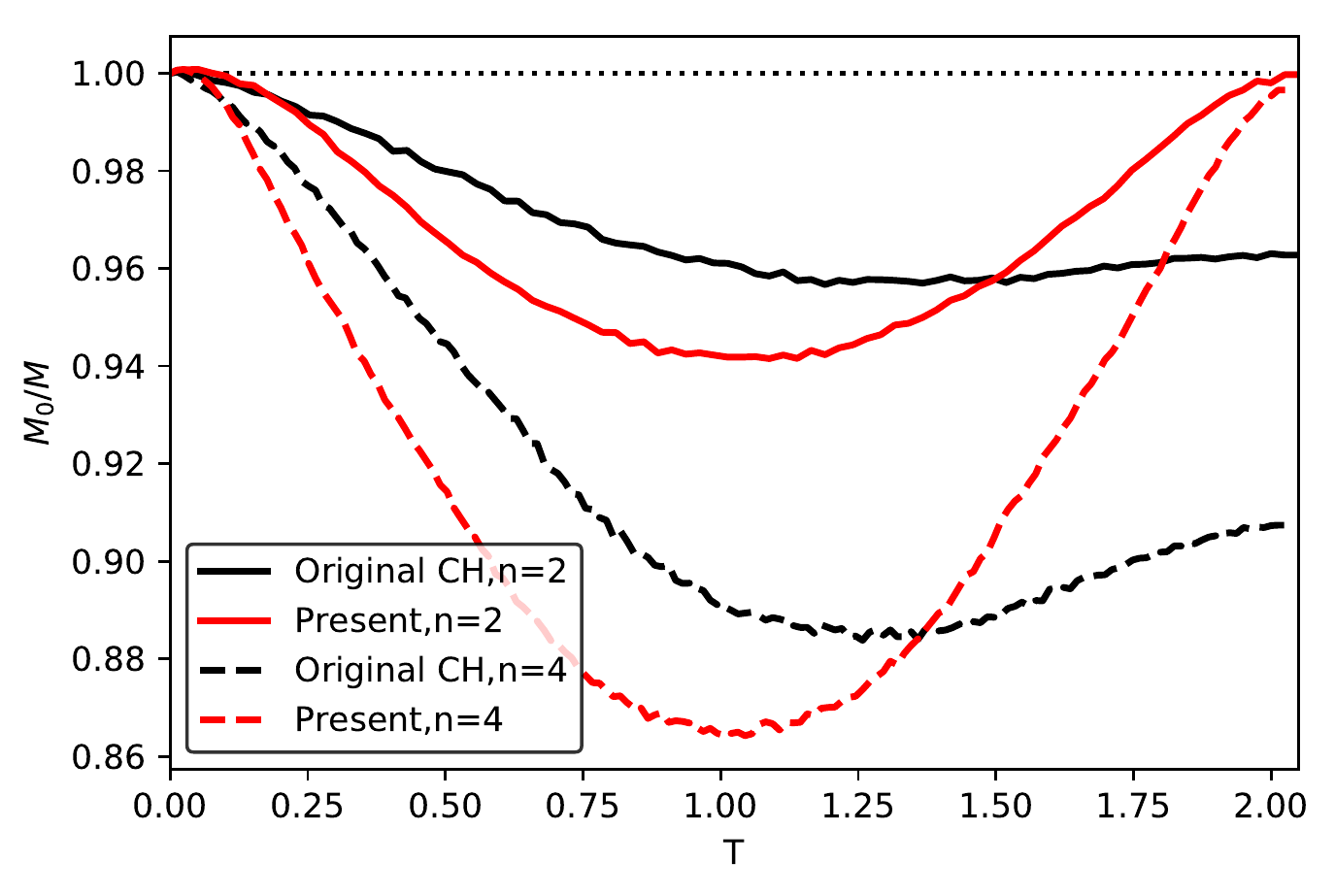}}
\caption{ Temporal evolution of the normalized area of the circle.}
\label{fig:vortex_compareMass}
\end{figure}

In order to validate the proposed method the performance of tracking the interface, the problem of Rider and kothe (1995) is simulated in a square domain $L\times L$~\cite{balcazar2014finite}. A circular droplet with radius $R$ initially is placed at $(0.5L,0.75L)$.
The flow field is given as~\cite{balcazar2014finite,liang2014phase},
\begin{equation}
\label{eq:velocity}
\begin{aligned}
u=U_0\sin^2(\pi x)\sin(2\pi y)\cos\left(\frac{\pi t}{T}\right),\\
v=-U_0\sin^2(\pi y)\sin(2\pi x)\cos\left(\frac{\pi t}{T}\right),
\end{aligned}
\end{equation}
where $T=nL/U_0$ with $n$ being a positive integer. Based on Eq.~(\ref{eq:velocity}),  the circular droplet  will undergo the largest deformation at $T$ and  come back to its initial position at $2T$.
In the simulations, the parameters are set as $L=400, \sigma=0.001, W=4, \tau_g=0.5+\sqrt{3}/6, \eta=2, U_0=0.02, \lambda=0.05$.
The evolution of  the interface advection and deformation obtained by the present method is shown in Fig.~\ref{fig:vortex}. It can be found that the interface is stretched into a very long filament that spirals around the center of the domain, which compares very well with that obtained using other high-order methods~\cite{balcazar2014finite,sheu2009development}. In particular, the values of the order parameter stay in the physically reasonable interval [-1,1] in the whole process.
Fig.~\ref{fig:vortex_compareShapeHalf} shows the result of both the present method and the original CH method at $T$. It can be seen that the interface shapes predicted by both methods are significantly different at the tail. However, the results of the present method are consistent with those in \cite{desjardins2008accurate}. This is because the present method attempts to maintain the volume conservation, which leads to the formation of drops of the size of a few grids.
Fig.~\ref{fig:vortex_compareShapeFinal} shows the initial and the final profile of the interface advected to periods $2T$ with $n=2, 4$.
From Fig.~\ref{fig:vortex_compareShapeFinal}, we can see that the shapes of the droplet predicted by both methods can  return to its original position at $2T$.
However, the interface profile predicted by the present method returns to its original location of the interface with greater accuracy than the one of the original CH method.
The wiggles appearing are mainly due to the stringent velocity field and rapid shift in its direction at $T/2$ and $3/2T$.
Finally, we compare the evolution of the normalized area of the circle over time for both methods, as shown in Fig.~\ref{fig:vortex_compareMass}. At $2T$, most of the area of the droplet  in the present method can be recovered and the maximum volume loss  of the circular droplet is $0.346\%$. In comparison,  the area of the droplet in the original CH method has significant volume loss and  the maximum volume loss  of the circular droplet is $9.26\%$.

\subsection{Effect of parameter $\lambda$}
\begin{figure}[H]
\centering
\subfloat[]{\includegraphics[width=0.5\textwidth]{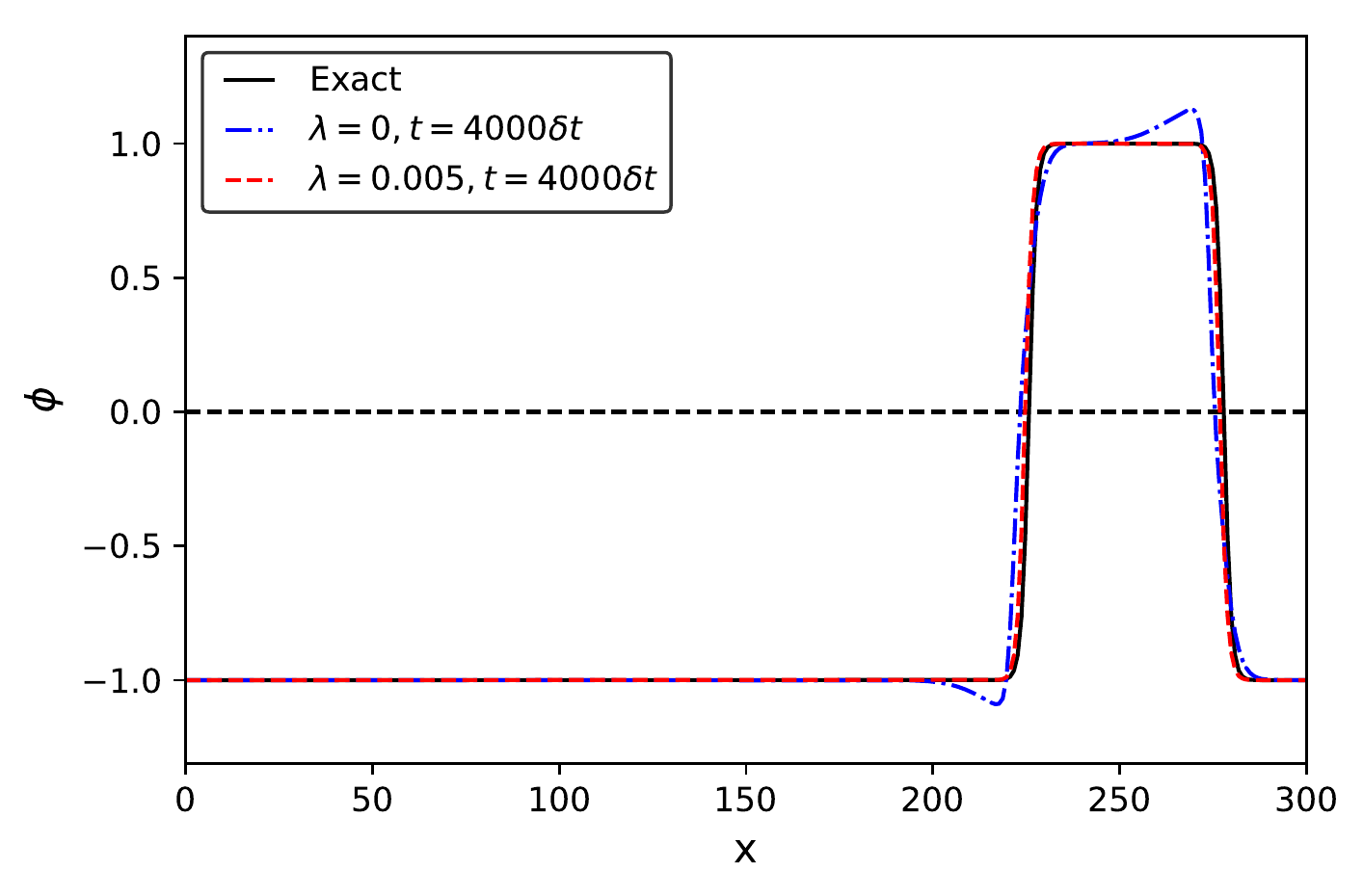}}~
\subfloat[]{\includegraphics[width=0.5\textwidth]{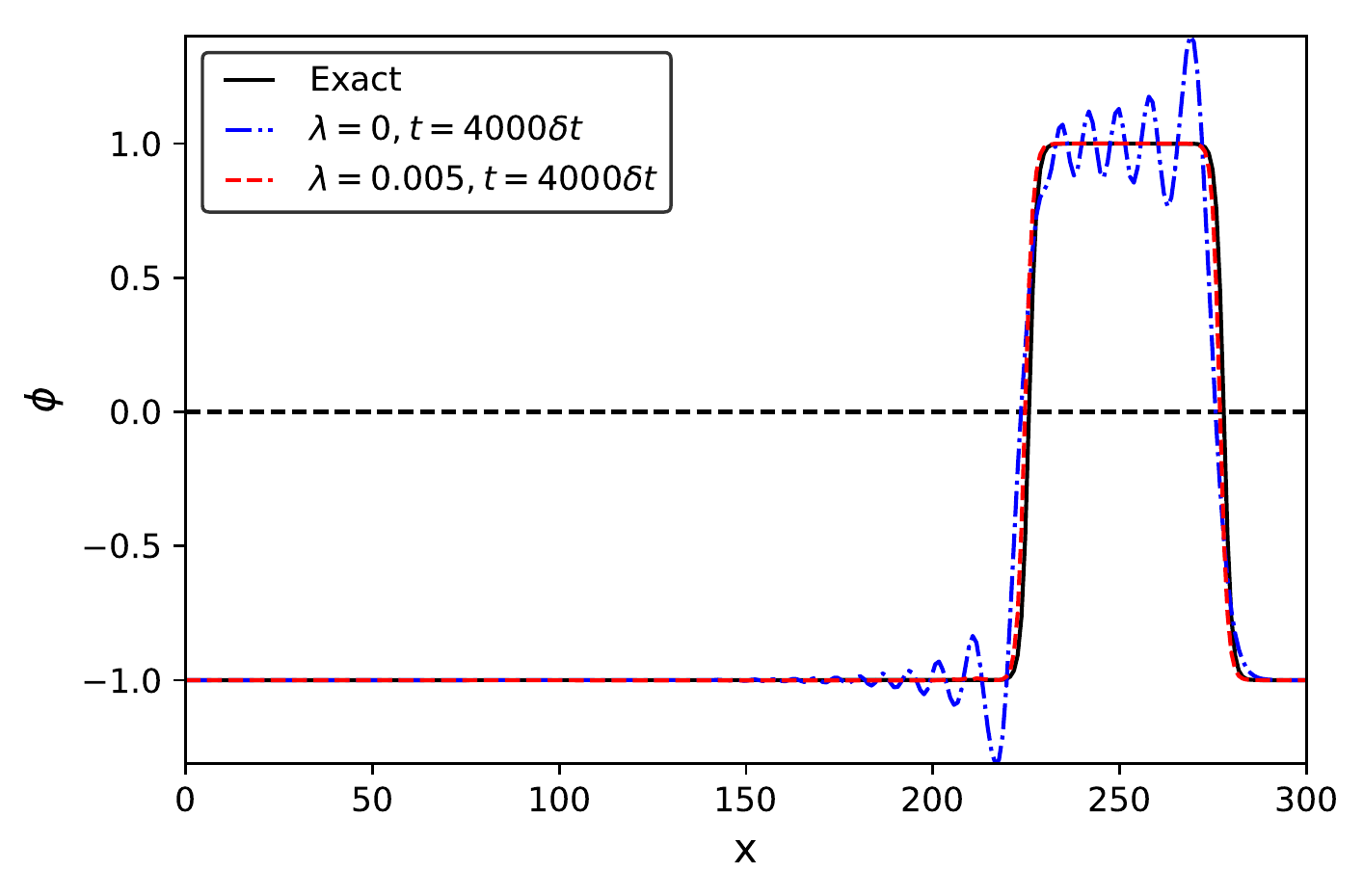}}\\
\caption{Translation of a circular drop with (a) $\sigma=0.01$ and (b) $\sigma=0.0001$}
\label{fig:compare}
\end{figure}

To investigate the effect of parameter $\lambda $ in Eq.~(\ref{eq:final_mCHE}) on the interface between two phases, we consider a translation of a circular drop of radius $R$  with a constant velocity on the domain $L\times 3L$. The initial shape is set as
\begin{equation}\label{eq:lambda_effect_initial}
\phi(x,y)=\tanh\left(2\frac{R-\sqrt{(x-50)^2+(y-50)^2}}{W}\right).
\end{equation}
The exact solution at time t can be obtained by $\phi(x,y)=\tanh(\frac{2(R-\sqrt{(x-50-ut)^2+(y-50)^2})}{W})$.
In the simulations, the velocity is set as $\bm u=(u,v)=(0.01,0)$.
The other parameters are given by $L=100$, $\tau_g=0.5+\sqrt{3}/6$, $W=4$, $R=26$.
 We run the simulation up to $t=4000\delta t$. Comparisons between the numerical results of  the modified CH equation with different values of $\lambda$ are shown in Fig.~\ref{fig:compare}.
It can be found that the values of $\phi$ obtained by the original CH method ($\lambda=0$) are slightly smaller than $-1$ on the left and larger than $1$ on the right. By contrast, the interfaces shape obtained by the present model with $\lambda=0.05$ are in excellent agreement with the exact solutions.
 In addition,  the location of the interface predicted by the original CH method deviates from that of the exact solution  while the present method can track the interface location correctly.

To furthermore quantitatively investigate the effect of $\lambda$ on the results, we define the relative maximum and minimum errors of the order parameter as follows
\begin{equation}\label{eq:maxminPHI}
 \phi_{max}=\text{max}\left|\frac{\phi(x,y)-1}{2}\right|, \qquad \phi_{min}=\text{max}\left|\frac{\phi(x,y)+1}{2}\right|.
\end{equation}
Then, we repeat the above test  with $\lambda=0, 0.001,0.005,0.01,0.05$. The relative maximum and minimum errors of the order parameter are presented in Tables.~\ref{fig:maxmum_Errors} and ~\ref{fig:minimum_Errors}.
 It can be found that the relative errors decrease significantly as $\lambda$ increases, which means that the numerical values of the order parameter are gradually adjusted to $-1$ or $1$. However, for larger $\lambda$, relative maximum and minimum errors of the order parameter cannot continue to decrease. A tiny deviation still exists. In addition, a larger $\lambda$ may give rise to numerical instability. Therefore, for the rest of the paper, we will use $\lambda=0.005 $ for presentation unless otherwise stated.
\begin{table}[!htb]
  \centering
  \caption{The Relative maximum errors with various $\lambda$.}\label{fig:maxmum_Errors}
\setlength{\tabcolsep}{4mm}{%
\begin{tabular}{lccccccc}
\hline
\multirow{2}*{$\lambda$}& \multicolumn{3}{c}{$U=0.01$}& & \multicolumn{3}{c}{ $U=0.05$}\\
\cline{2-4} \cline{6-8}
  & $\sigma=0.01$ &$\sigma=0.001$& $\sigma=0.0001$&   &$\sigma=0.01$& $\sigma=0.001$&$\sigma=0.0001$\\
\hline
$0.0 $      & 0.06314         &0.13071           &0.20146            &
                  & 0.10459         &0.19082           &0.21659            \\
$0.001$     &0.01638          &0.01384           &0.01708             &
                  &0.06907          &0.09880           &0.10610            \\
$0.005$     &0.00002          &0                 &0.00254            &
                  &0.01356          &0.01665           &0.01805            \\
$0.01$      &0                &0.00004           &0.00168            &
                  &0.00005          &0.00058           &0.00075            \\
$0.05 $     &0                &--                &--                 &
                  &0                & 0.00088          &0.00385            \\
\hline
\end{tabular}}
\end{table}
\begin{table}[!htb]
  \centering
  \caption{The Relative minimum errors with various $\lambda$. }\label{fig:minimum_Errors}
\setlength{\tabcolsep}{4mm}{%
\begin{tabular}{cccccccc}
\hline
\multirow{2}*{$\lambda$}& \multicolumn{3}{c}{$U=0.01$}& & \multicolumn{3}{c}{ $U=0.05$}\\
\cline{2-4} \cline{6-8}
 & $\sigma=0.01$ &$\sigma=0.001$& $\sigma=0.0001$& &$\sigma=0.01$& $\sigma=0.001$&$\sigma=0.0001$\\
\hline
$0.0$       &0.04495      &0.10503        &0.15960       &
                  &0.08267      &0.15024        &0.17401       \\
$0.001$     &0.01276      &0.01210        &0.01552       &
                  &0.05657      &0.08323        &0.09033      \\
$0.005$     &0.00002      &0              &0.00274       &
                  &0.01183      &0.01506        &0.01631             \\
$0.01$      &0            &0.00006        &0.00166       &
                  &0.00005      &0.00053        &0.00070             \\
$0.05$      &0            &--             &--            &
                  &0.00002      &0.00072        &0.03209             \\
\hline
\end{tabular}}
\end{table}

\subsection{Laplace-Young law}
\begin{figure}[H]
\centering
\subfloat[]{\includegraphics[width=0.5\textwidth]{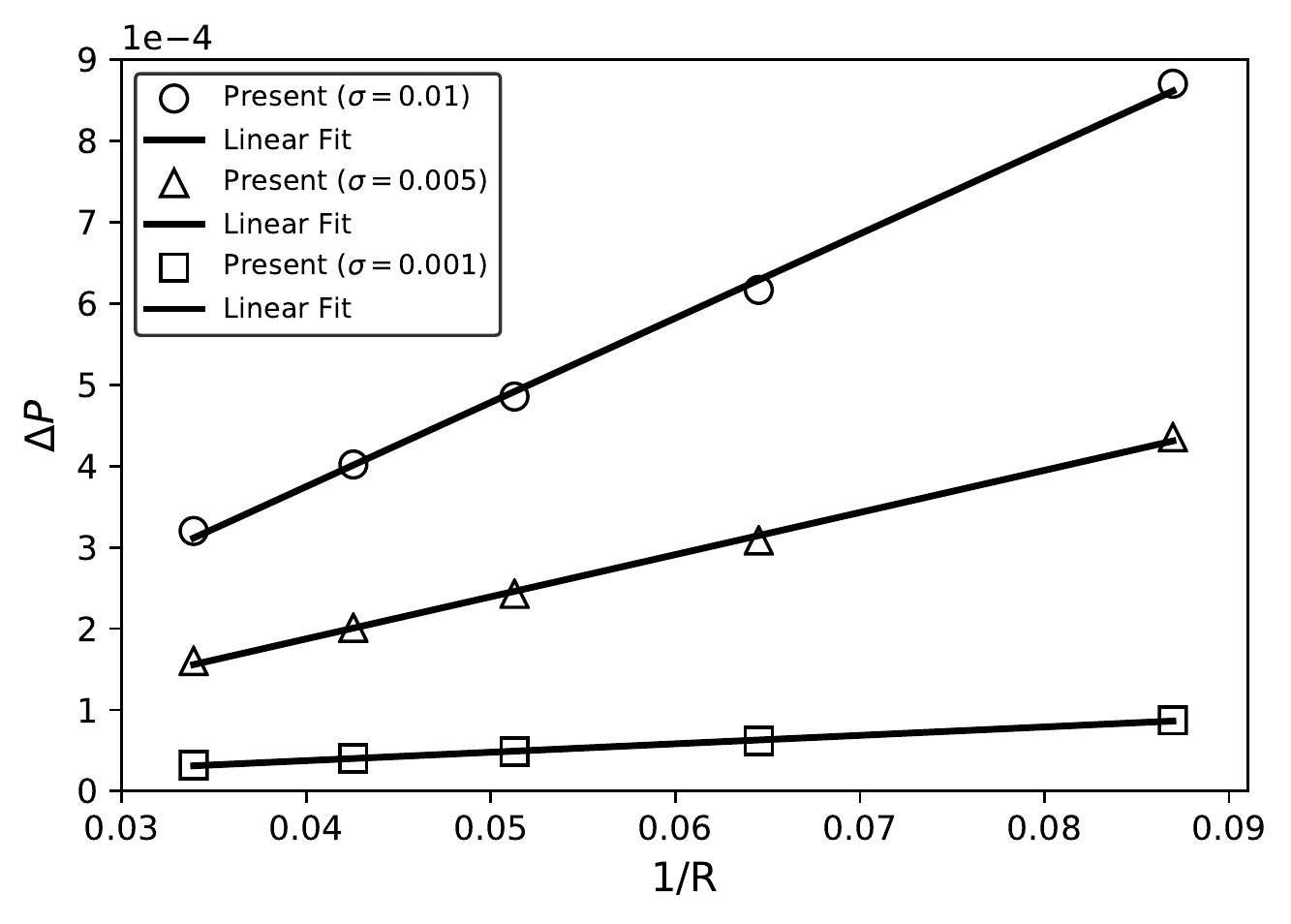}}
\caption{Laplace-Young law with $\rho_1/\rho_2=1000$ and $\mu_1/\mu_2=100$.}.
\label{fig:laplace}
\end{figure}

All of the tests are carried out by a given velocity field and the hydrodynamic equation is not considered.
To validate the accuracy of the proposed model coupled with the hydrodynamic equation, we carry out the Laplace-Young law test. Initially, a static circular droplet with radius $R$ is located in the center $(x_c,y_c)$ of a domain size $L\times L$. Periodic boundary conditions are applied on all the boundaries, and the order parameter is given by
\begin{equation}\label{eq:case1_intial_phi}
  \phi(x,y)=\tanh\left(\frac{2(R-\sqrt{(x-x_c)^2+(y-y_c)^2})}{W}\right).
\end{equation}
 Theoretically, in the equilibrium state, the pressure difference ($\Delta p$) between the inside and outside droplet is equivalent to the ratio of the surface tension and the radius  of the droplet, i.e., $\Delta P=\sigma/R$.
 The density and dynamic viscosity of the droplet are fixed as $\rho_1=1000, \rho_2=1, \mu_1=10, \mu_2=0.1$. In Eq.~(\ref{eq:filter}), the coefficients of ten-order filter (SF-11) are used and the intensity of the filter is fixed at $a=0.5$ unless otherwise stated.
And the other parameters are taken as: $L=100,  x_c=y_c=50, \lambda=0.005, \tau_g=1/2+\sqrt{3}/6, \eta=2$ and $W=4$.
Fig.~\ref{fig:laplace} shows the relationship between the pressure differences  and the reciprocal of different radii of the droplet with three surface tensions ($\sigma=0.01, 0.005, 0.001$) measured after $2\times10^5\delta t$.  The pressure differences in Fig.~\ref{fig:laplace} show good agreement with the Laplace-Law, which verifies the accuracy of the proposed model in modeling the interface tension with large density ratios.

In addition,  the volume of the droplet with different radius (R=10, 20, 30, 40) are measured to validate the volume-conserved property for the present method. To quantitative analysis,  the percentage error in the droplet volume is calculated by $\text{Err}=((\int_{\phi(\bm x,0)>0} 1d\bm x -\int_{\phi(\bm x,t)>0}1d\bm x)/\int_{\phi(\bm x,0)>0} 1d\bm x )\times 100$,
where $\phi(\bm x,0)$ and $\phi(\bm x, t)$ are the order parameter at $t=0$ and $t=2\times 10^5\delta t$, respectively.
As the numerical instability occurs for the original CH method with large density contrast, the parameters in this case are set as, $L=200, \rho_1=10, \rho_2=1,\sigma=0.001, \tau_f=0.6$.
The results are summarized in Table \ref{tab:mass}. It can be observed that the volume loss of the droplet obtained by the original CH method increases with the decreasing radius of the droplet while the volume of the droplet obtained by the present method conserve volume exactly.
\begin{table}[h]
  \centering
  \caption{Relative errors in the volume of a droplet with different radii at $t=2\times10^5\delta t$.}\label{tab:mass}
\begin{tabular}{p{3cm}p{2cm}p{2cm}p{2cm}p{2cm}}
\hline
  Raidus& 10 & 20&30 &40\\
  \hline
 Original CH& 3.9344 &0.6426 & 0.2848&0.1596  \\
  Present & 0& 0  &0 & 0\\
  \hline
\end{tabular}
\end{table}

\subsection{ Capillary wave}
\begin{figure}[H]
\centering
\subfloat{\includegraphics[width=0.45\textwidth]{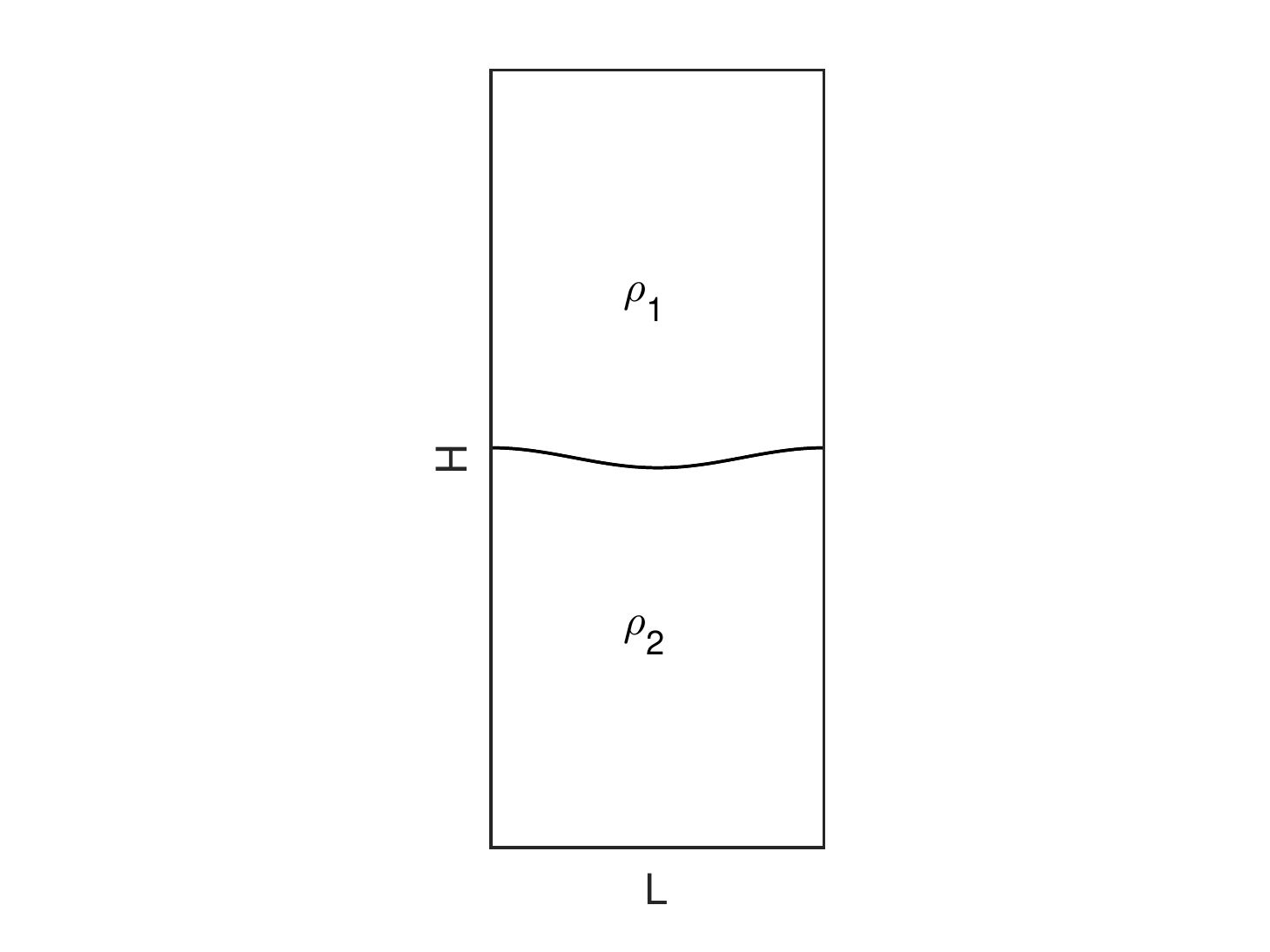}}
\caption{Schematic illustration of the initial condition for the capillary wave problem.}
\label{fig:CapillaryInitial}
\end{figure}

\begin{figure}[H]
\centering
\subfloat[]{\includegraphics[width=0.45\textwidth]{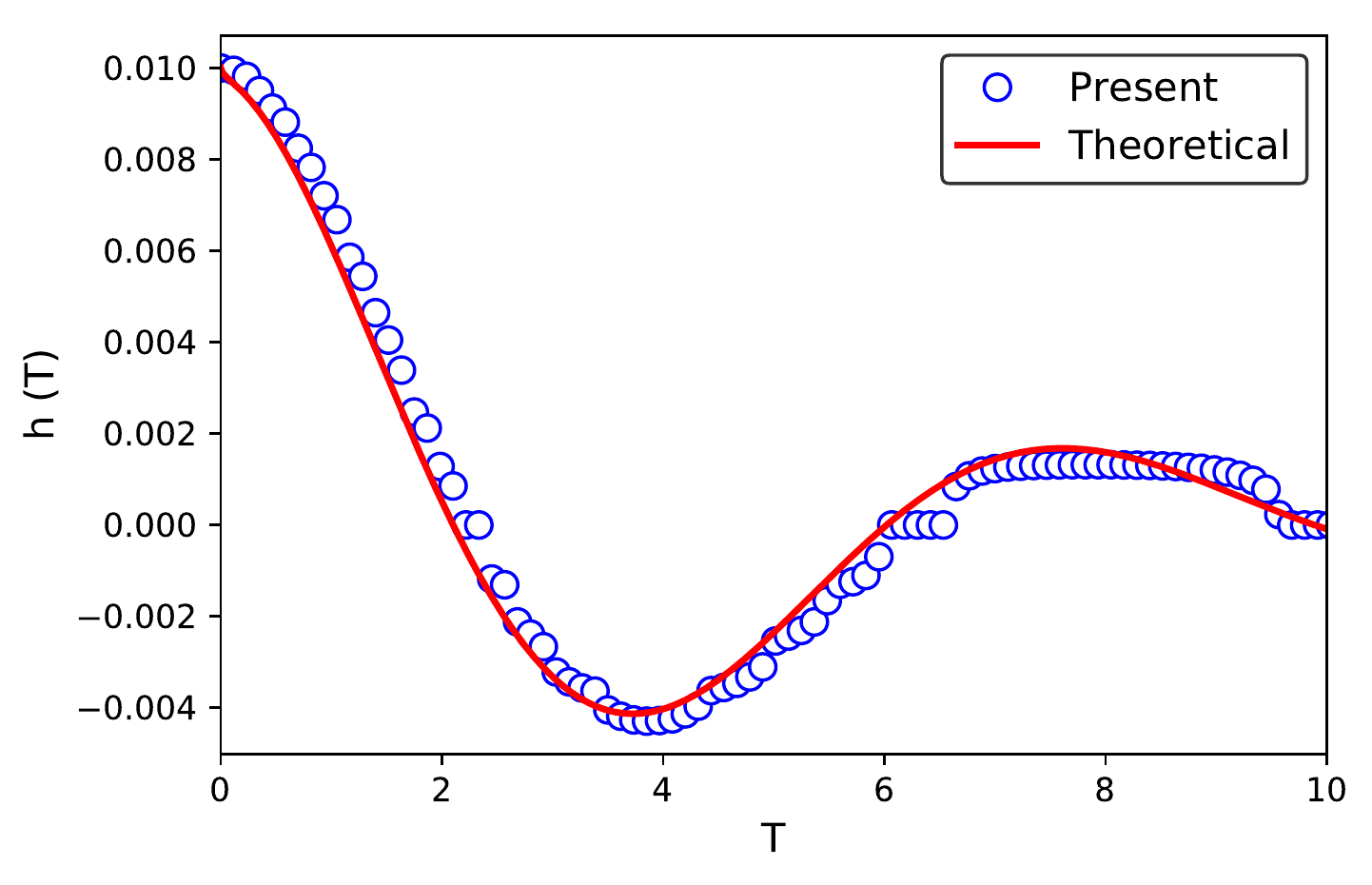}}~
\subfloat[]{\includegraphics[width=0.45\textwidth]{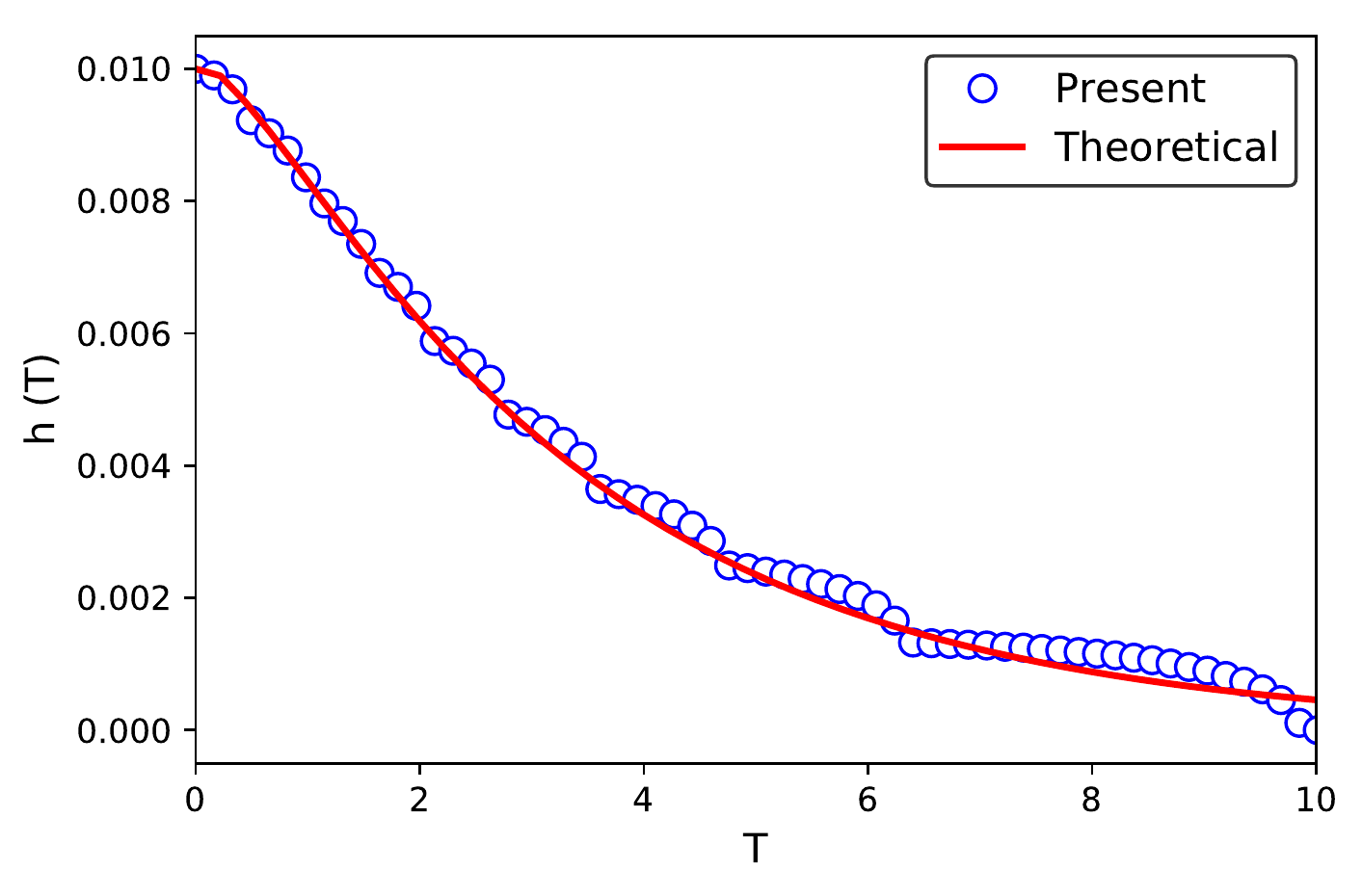}}\\
\subfloat[]{\includegraphics[width=0.5\textwidth]{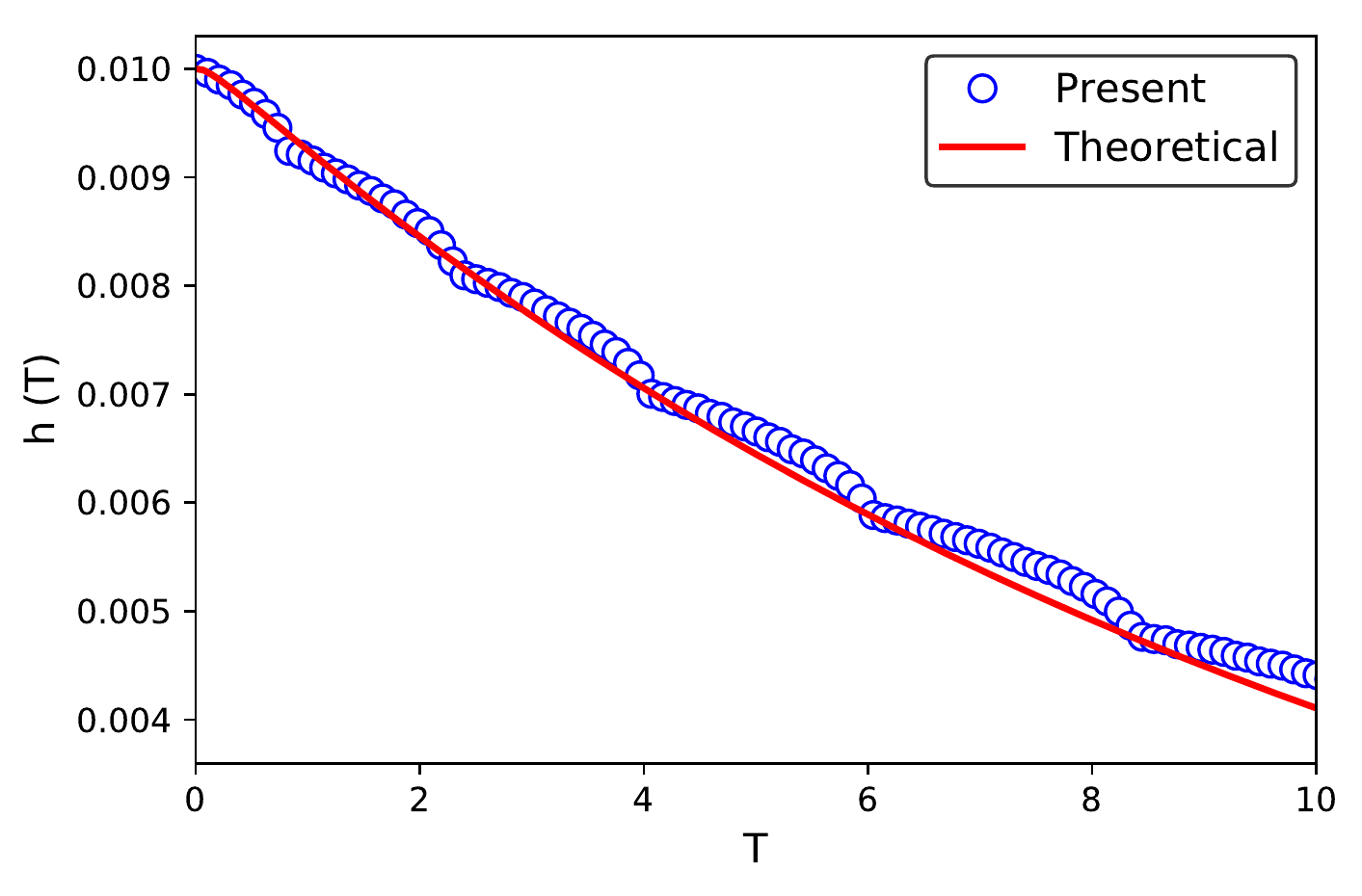}}
\caption{Evolution of wave amplitude with (a) $\rho_2/\rho_1=1$, (b) $\rho_2/\rho_1=100$, (c) $\rho_2/\rho_1=1000$. The time is normalized by the characteristic time $t\omega_0$.}
\label{fig:Capillary}
\end{figure}

To further validate the present method, a capillary wave problem~\cite{prosperetti1981motion,shao2015hybrid} is carried out. As shown in Fig.~\ref{fig:CapillaryInitial},
a heavier fluid (labeled as fluid 2) is placed under a lighter fluid (labeled as fluid 1) with  a small perturbation $y=H/2+ H_0 \cos(\kappa x)$ on the interface, where $H$ is the height of the computation domain, $H_0$ is the initial perturbation amplitude and $\kappa$ is the wave number. Under the influence of surface tension and viscosity, the capillary  wave  with a decaying amplitude oscillates until the fluid is at rest.
According to the work of Prosperetti~\cite{prosperetti1981motion}, the evolution of interface wave amplitude $h(t)$ is given by
\begin{equation}\label{capillary_wave}
\frac{h(t)}{H_0}=\frac{4(1-4\beta)\nu^2\kappa^4}{8(1-4\beta)\nu^2\kappa^4+\omega_0^2}\mbox{erfc}(\sqrt{\nu\kappa^2 t})+
\sum_{i=1}^{4} \frac{z_i}{\bm Z_i}\frac{\omega_0^2}{z_i^2- \nu \kappa^2}e^{( z_i^2-\nu \kappa^2)t} \mbox{erfc}(z_i \sqrt{t}),
\end{equation}
where $\beta$ is defined as $\rho_1 \rho_2/(\rho_1+\rho_2)^2$, $\nu$ is the kinematic viscosity, $\omega_0^2=(\sigma \kappa^3)/(\rho_2+\rho_1)$ and $\mbox{erfc}(z_i)$ is the complementary error function of complex $z_i$, which is the solution of the following equation
\begin{equation}\label{error_function}
z^4-4\beta \sqrt{\nu \kappa^2} z^3+2(1-6\beta)\nu \kappa^2 z^2+4(1-3\beta)(\nu \kappa^2)^{3/2}z+(1-4\beta)\nu^2\kappa^4+\omega_0^2=0,
\end{equation}
and $ Z_i$ is defined as
\begin{equation}\label{z}
 Z_i=\prod_{j=1,j\neq i}^{j=4}(z_j-z_i),i=1,\cdots,4.
\end{equation}
For this case, the computation domain is set as $L\times H=1\times 3$, the kinetic viscosity is $0.01$ and the surface tension is $0.02$. The initial amplitude of the perturbation wave is $H_0=0.01H$. The wave number is set as $\kappa=2\pi/L$. No-slip boundary conditions are imposed on the top and bottom wall while periodic boundary conditions on the side directions. In the simulations, a grid size of $300\times 900$ is applied to discretize the computational domain. The other parameters are set as  $\tau_g=0.5+\sqrt{3}/6$, $W=4$, $\lambda=0.005$. In addition, the non-dimensional time is normalized by $T=t\omega_0$.
Fig.~\ref{fig:Capillary} shows the comparison of  the numerical results and the theoretical predictions for $\rho_2/\rho_1=1, 100, 1000$. It can be seen that the numerical results are in good agreements with the theoretical solution. However, the attenuation of the amplitude in the theoretical is faster than the one predicted from the simulation. This delay has been pointed out in \cite{denner2016frequency} that the nonlinear effects are negligible due to the linearized Navier-Stokes equations assumption. Overall, this validates the accuracy of the proposed model in capturing the interface of immiscible two-phase fluids.

\subsection{Rising bubble}
\begin{figure}
\centering
\subfloat{\includegraphics[width=0.25\textwidth]{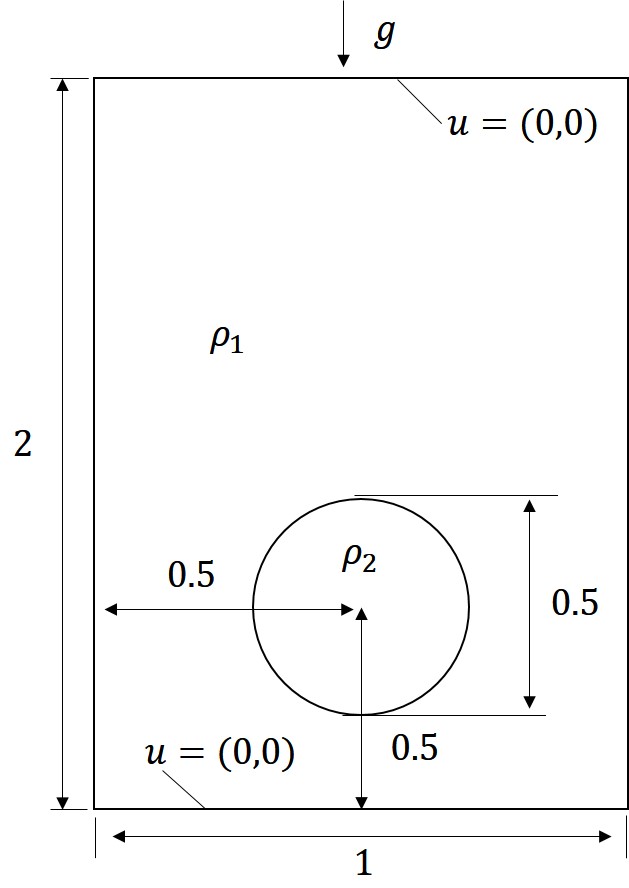}}
\caption{Initial configuration for the rising bubble}
\label{fig:Rising_initial}
\end{figure}

In this section, we examine the present model by simulating a single bubble rising in a liquid column and compare the results with the data in the literature. For this problem, a numerical benchmark configuration has been proposed for two-dimensional bubble dynamics, including a small density and viscosity ratio case (labeled as test 1) and a large density and viscosity ratio case (labeled as test 2)~\cite{hysing2009quantitative,aland2012benchmark}. Different numerical approaches have been quantitatively compared, such as Eulerian level set finite element methods, arbitrary Lagrangian-Eulerian moving grid approaches, and diffuse interface methods.
The schematic diagram of this problem is depicted in Fig.~\ref{fig:Rising_initial}. A bubble of radius $R$ is placed at $\text{(0.5m,0.5m)}$ in a rectangular domain $1 \text{m}\times 2 \text{m}$. The hydrodynamics are governed by two non-dimensional  numbers: the Reynolds number ($\text{Re}$) and the Eotvos number (Eo),
\begin{equation}\label{eq:case2-dimensionless}
  Re=\frac{\rho_1 U_g 2R}{\mu_1},\qquad Eo=\frac{2\rho_1 U_g^2 R}{\sigma},
\end{equation}
where $g$ is the gravitational acceleration, $U_g=\sqrt{g2R}$ is the gravitational velocity. No-slip conditions are applied on the upper and lower walls and periodic conditions are used on the left and right boundaries. In order to quantify the dynamics of the bubble during its course of rise,
the following benchmark quantities are measured, including rise velocity and center of mass, which are defined as,
\begin{equation}\label{eq:quantify-yc}
  y_c=\frac{\int_{\phi<0}y dx}{\int_{\phi<0} 1 dx},
\end{equation}
\begin{equation}\label{eq:quantify-uc}
  V_c=\frac{\int_{\phi<0}v dx}{\int_{\phi<0} 1 dx},
\end{equation}
where $v$ is the velocity component in the vertical direction.
The fluid and physical parameters for both tests are listed in table ~\ref{tab:test_Paras}.
In the simulations, a grid size of $240 \times 480$ is applied to discretize the flow domain. The gravitational acceleration and surface tension are determined by the non-dimensional numbers.
\begin{table}
\centering
\caption{Physical parameters and dimensionless numbers}
\label{tab:test_Paras}
\begin{tabular}{ccccccccc}%
\hline
 Test &Eo & Re & $U_g$ & $\rho_1$ &  $\rho_2$ & $\mu_1$ & $\mu_2$ \\
\hline
1&10  & 35&  0.0024 & 1000& 100& 10&   1 \\
2&125 & 35&  0.0019 & 1000&   1& 10& 0.1\\
\hline
\end{tabular}
\end{table}

\begin{figure}
\centering
\includegraphics[width=0.75\textwidth]{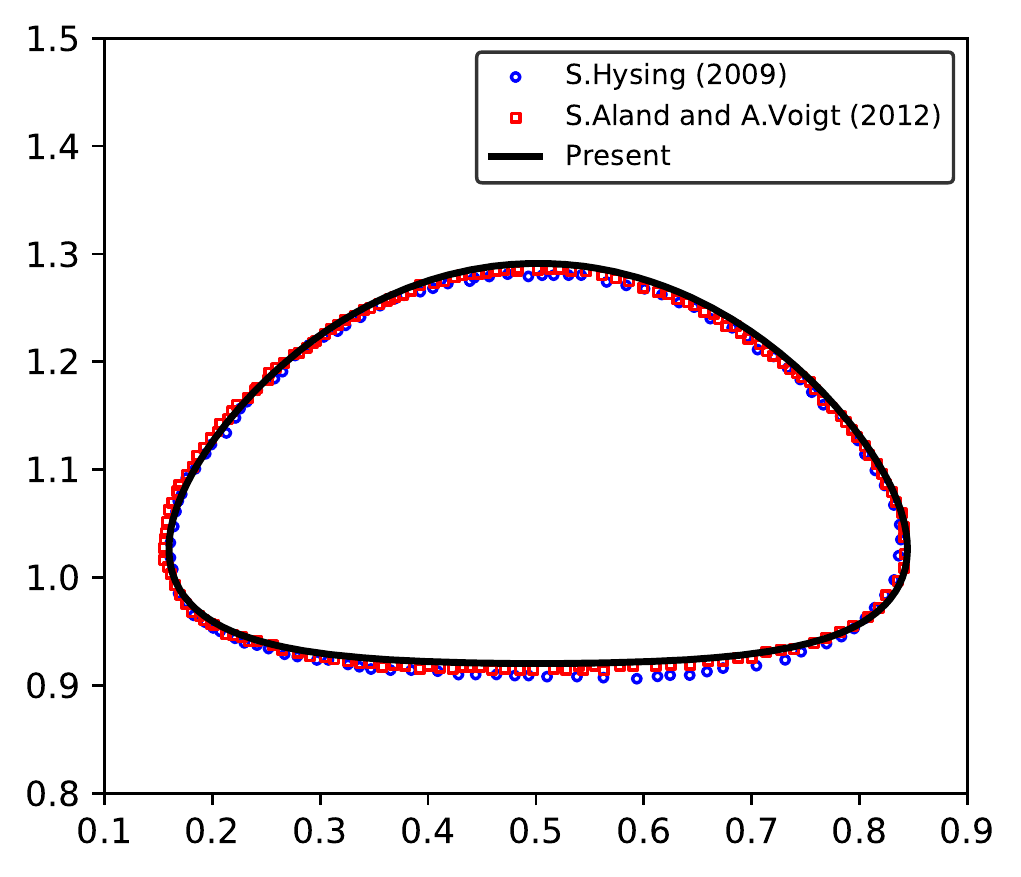}
\caption{Test 1: bubble shapes at $t=3$.}
\label{fig:test1-shape}
\end{figure}

\begin{figure}
\centering
\includegraphics[width=0.5\textwidth]{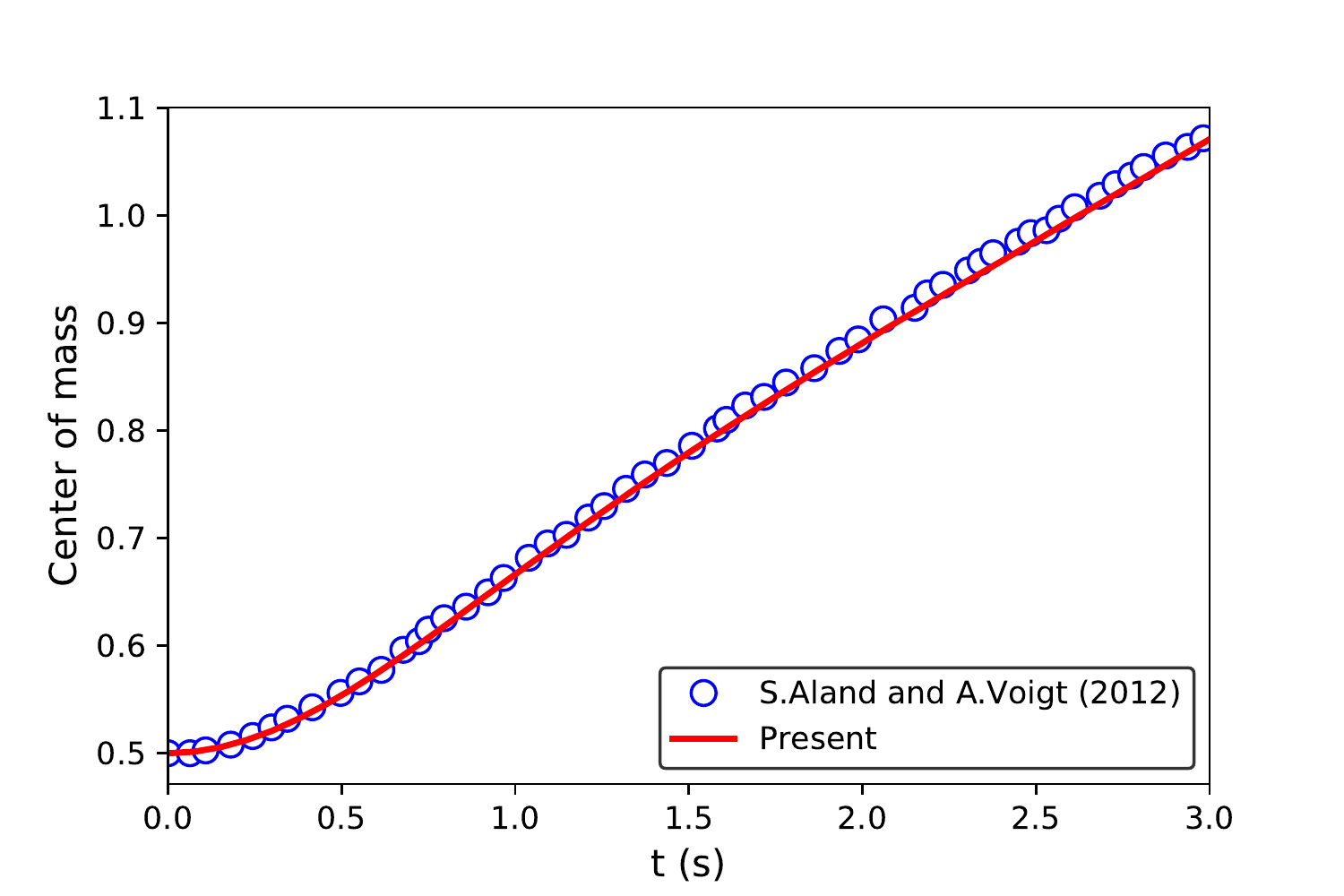}~
\includegraphics[width=0.5\textwidth]{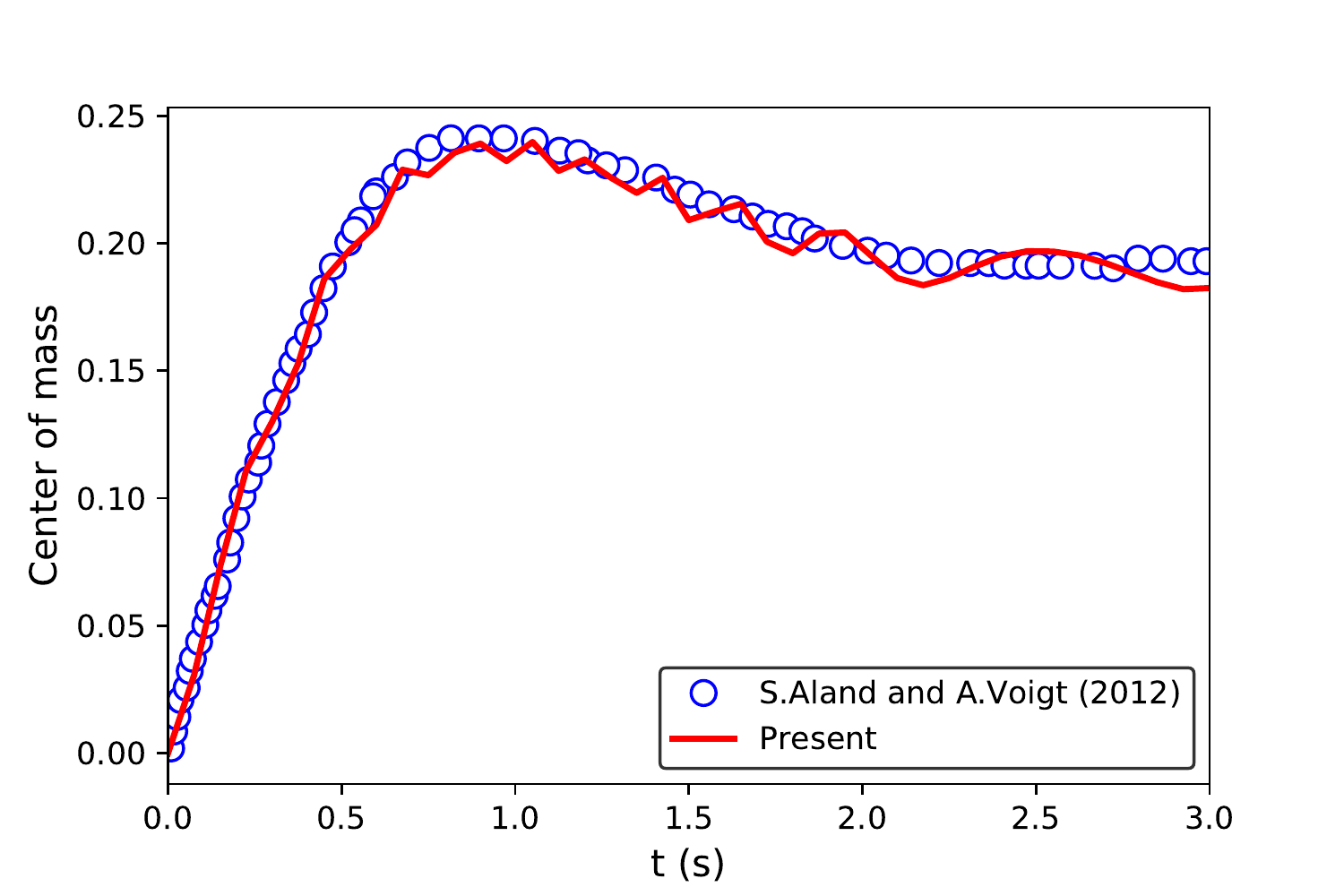}~
\caption{Test 1: (a) center of mass; (b) rise velocity.}
\label{fig:test1-mass-velocity}
\end{figure}

Fig.~\ref{fig:test1-shape} compares the bubble shapes predicted by the present method at the final time (t=3s) with the benchmark solutions of Aland and Voigt~\cite{aland2012benchmark} in terms of test 1.
It can be observed that the overall interface shapes predicted by the present method agree well with those of published data. Furthermore, Fig.~\ref{fig:test1-mass-velocity} compares the center of mass and the rising velocity for the rising bubble with the benchmark solutions of Aland and Voigt~\cite{aland2012benchmark}. It can be found that the results obtained by the present method are almost in consistent with the results from the literature.

\begin{figure}
\centering
\includegraphics[width=0.75\textwidth]{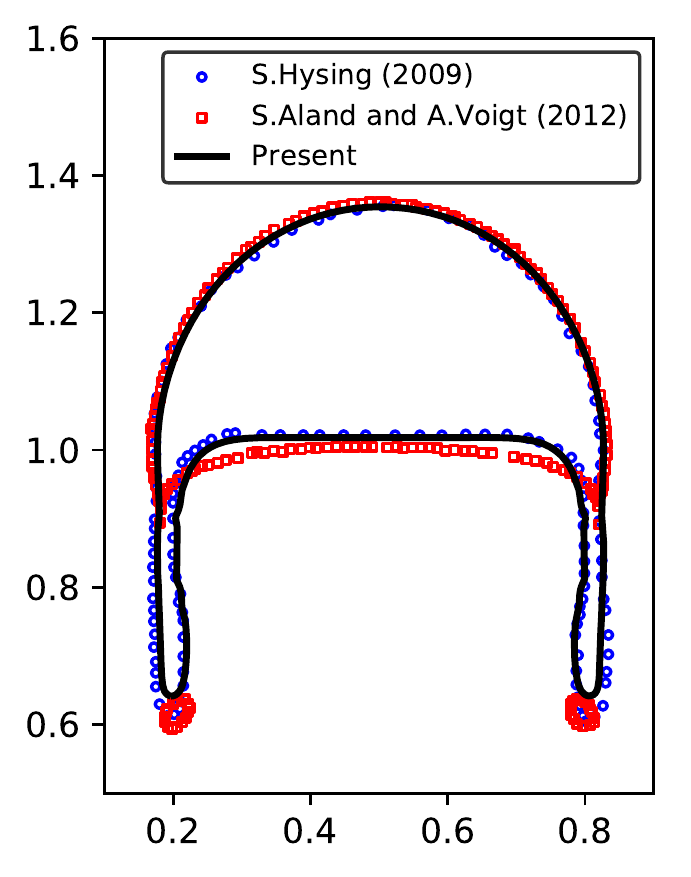}
\caption{Test 2: bubble shapes at $t=3$ .}
\label{fig:test2-shape}
\end{figure}

\begin{figure}
\centering
\includegraphics[width=0.5\textwidth]{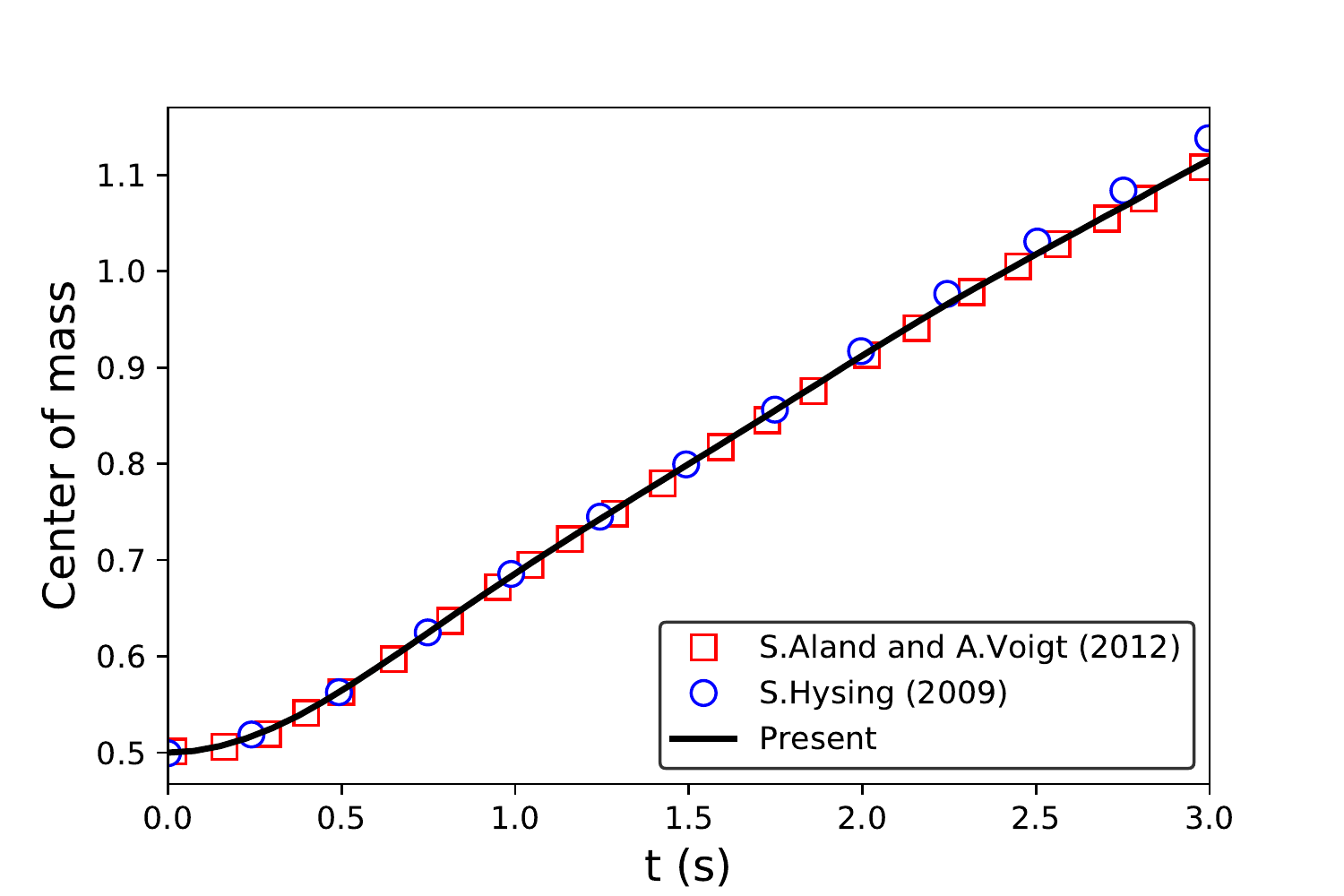}~
\includegraphics[width=0.5\textwidth]{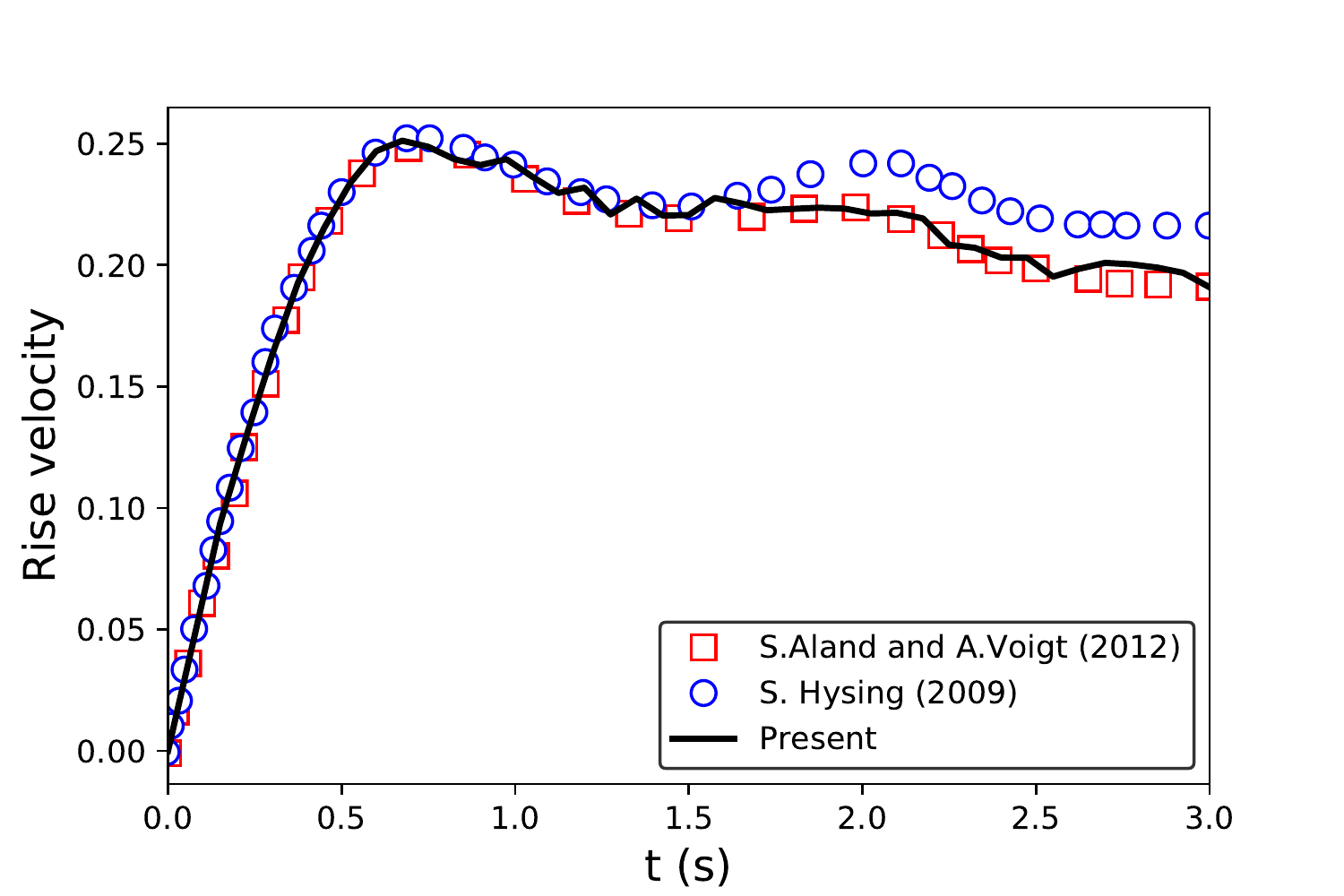}
\caption{Test 2: (a) center of mass; (b) rise velocity.}
\label{fig:test2-mass-velocity}
\end{figure}

 We also carried out the simulation with a density ratio 1000 and viscosity ratio of 100 in terms of test 2. Both the bubble shapes predicted by the present method and those of published data at $t=3s$ are shown in Fig.~\ref{fig:test2-shape}. From  Fig.~\ref{fig:test2-shape}, the shape of bubble experiences significant topology change. And there are visible differences between all of the three models. The thin filaments break up for the results of Hysing \emph{et al}~\cite{hysing2009quantitative} while no break off occurs for both the results of S. Aland and A. Voigit~\emph{et al.}\cite{aland2012benchmark} and the present method. However, the overall interface shapes and positions are quite similar. This difference may be caused by the grid resolution because a appropriate thickness of the interface is required for the diffuse interface approximations. As the interface thickness tends to zero, the differences between the results of the three methods should be rather small.
 In addition, the center of mass and rise velocity over time are shown in Fig.~\ref{fig:test2-mass-velocity}. Again, good agreements with the published results are demonstrated. Based on the above results, the accuracy and performance of the present method are confirmed.

\section{CONCLUSIONS}
In this paper, a MRT-LBE method for multiphase flow with large density ratios is developed based on the phase-field theory. To accurately capture the interface, the modified CH equation with the interface correction term and the flux correction term is solved by a fractional scheme. In addition, a high-order compact selective filter is applied to macroscopic quantities (pressure and velocity) to improve the numerical stability.
Thereafter, several numerical benchmarks for two-phase flows are performed to evaluate the accuracy and stability of the proposed method.
First, the results of both the single vortex deform of a circular droplet and the translation of a drop have been compared fairly well with the theoretical interface shape and location. The accuracy of the proposed method for capturing the interface is demonstrated. By coupling the hydrodynamic equations, the present method is further assessed by the Laplace-Young law and the capillary wave problems. The results are in good agreement with the theoretical solutions.
Finally, a rising bubble with large density ratios are simulated to examine the performance of the present method, which demonstrates the capability of the present method for problems with complex interface deformation  and large density ratios.

\section*{ACKNOWLEDGEMENTS}
\begin{appendix}
\section{CHAPMAN-ENSKOG ANALYSIS OF THE MRT-LBE MODE FOR THE NAVER-STOKES EQUATIONS}
\label{eq:derivationNS}
To recover the incompressible immiscible two-phase governing equations from Eq.~(\ref{eq:distribution_NS}), the following multi-scale expansions are introduced,
\begin{equation}\label{eq:expend}
\begin{aligned}
g_i&=g^{(0)}_i+\epsilon g^{(1)}_i+\epsilon^2 g^{(2)}_i+\ldots,\\
\partial_t &=\epsilon\partial_{t_1}  + \epsilon^2 \partial_{t_2}, \hspace{5mm}\nabla=\epsilon \nabla_1,\hspace{5mm} \bm F=\epsilon \bm F{(1)}+\epsilon^2 \bm F^{(2)},
\end{aligned}
\end{equation}
where $\epsilon$ is a small expansion parameter. By applying the Taylor expansion to Eq.~(\ref{eq:distribution_NS}), one can obtain
\begin{equation}\label{eq:Tayforg}
D_i g_i+\frac{\delta t}{2}D_i^2 g_i+\ldots=-\frac{1}{\delta t}\left(\bm{\text{M}}^{-1}\bm{\text{S}}^g\bm{\text{M}}\right)_{ij}\left(g_j-g_j^{eq}\right)
+ \left[\bm{\text{M}}^{-1}
\left(\bm I-\frac{1}{2}\bm{\text{S}}^g\right)\bm{\text{M}}\right] G_j,
\end{equation}
where $D_i=\partial_t+c_i\cdot \nabla$. Substituting Eq.~(\ref{eq:expend}) into (\ref{eq:Tayforg}) and taking the zeroth, first-, and second-order in $\epsilon$ leads to
\begin{align}
\label{eq:g0}
\epsilon^0:\quad & g_i^{(0)}=g_i^{eq}\\
\label{eq:g1}
\epsilon^1:\quad& D_{1i} g_i^{(0)} = -\frac{1}{\delta t} \left(\bm{\text{M}^{-1}}\bm{\text{S}}^g\bm{\text{M}}\right)_{ij} g_j^{(1)}
+\left[\bm{\text{M}^{-1}}\left(\bm I-\frac{\bm{\text{S}}^g}{2}\right)\bm{\text{M}}\right]_{ij} G_j^{(1)},\\
\label{eq:g2}
\epsilon^2:\quad & \partial_{t_2} g_i^{(0)} + D_{1i} g_i^{(1)}+\frac{\delta t}{2}D_{1i}^2 g_i^{(0)}=-\frac{1}{\delta t}
\left(\bm{\text{M}^{-1}}\bm{\text{S}}^g\bm{\text{M}}\right)_{ij} g_j^{(2)}
+\left[\bm{\text{M}^{-1}}\left(\bm I
-\frac{\bm{\text{S}}^g}{2}\right)\bm{\text{M}}\right]_{ij} G_j^{(2)},
\end{align}
where $D_{1i}=\partial_{t_1}+\bm c_{i}\nabla$. Multiplying the matrix $\bm{\text{M}}$ on both sides of Eqs.~(\ref{eq:g0}-\ref{eq:g2}), one can obtain the equations in the moment space,
\begin{align}
\epsilon^0:\quad& \bm m_g^{(0)}=\bm m_g^{eq} \label{eq:Mg0}\\
\epsilon^1:\quad& \bm{\tilde{D}}_{1}\bm{m}_g^{(0)}=
-\bm{\text{S}}^g\bm m_g^{(1)} +
\left(\bm{\text{I}}-\frac{\bm{\text{S}}^g}{2}\right)\bm{\tilde{G}}^{(1)}\label{eq:Mg1}\\
\epsilon^2:\quad& \partial_{t_2} \bm m_g^{(0)} +\bm{\tilde{D}}_1 \left(\bm I-\frac{\bm{\text{S}}^g}{2}\right)\bm m_g^{(1)}
+\frac{\delta t}{2}\bm{\tilde{D}}_1 \left(\bm I- \frac{\bm{\text{S}}^g}{2}\right) \bm{\tilde{G}}^{(1)}
=-\bm{\text{S}}^g\bm m_g^{(2)}+
\left(\bm I-\frac{\bm{\text{S}}^g}{2}\right)\bm{\tilde{G}}^{(2)}\label{eq:Mg2},
\end{align}
where $\bm m_g=\bm{\text{M}}g$, $\bm{\tilde{G}}=\bm{\text{M}} G$, $\bm{\tilde{D}}_1=\bm{\text{M}}\bm D_1\bm{\text{M}}^{-1}$.  According to Eqs.~(\ref{eq:geq}) and (\ref{eq:Gi}), the elements of $\bm m_g$ and $\bm{\tilde{G}}$ can be expressed as,
\begin{equation}\label{mg0}
\bm m_g^{(0)}=\left(\frac{p}{c_s^2},m_{g_1}^{(0)},m_{g_2}^{(0)},
u_x,m_{g_4}^{(0)},u_y,m_{g_6}^{(0)},
m_{g_7}^{(0)},m_{g_8}^{(0)}\right)^T,
\end{equation}
\begin{equation}\label{mg1}
\bm m_g^{(1)}=\left(0,m_{g_1}^{(1)},m_{g_2}^{(1)},-\frac{F_x^{(1)}}{2}\delta t,m_{g_4}^{(1)},-\frac{F_y^{(1)}}{2}\delta t
,m_{g_6}^{(1)},m_{g_7}^{(1)},m_{g_8}^{(1)} \right)^T,
\end{equation}
\begin{equation}\label{mG1}
\begin{aligned}
\bm{\tilde{G}}_i&=\frac{1}{\rho}\left(0,6(\bm F-\nabla p)\cdot\bm u,-6(\bm F-\nabla p)\cdot\bm u,F_{x},-F_{x},F_{y},-F_y,\right. \\
&\left. 2(F_x-\partial_x p) u_x-2(F_y-\partial_y p) u_y,
(F_y-\partial_y p) u_x+(F_x-\partial_x p) u_y \right)^T,
\end{aligned}
\end{equation}
Substituting Eqs.~(\ref{mg0}-\ref{mG1}) into Eq.~(\ref{eq:Mg1}),
 the elements of Eq.~(\ref{eq:Mg1}) with $i=0,3,5$ can be written as,
\begin{equation}\label{eq:epsilon1-035-mac}
\begin{aligned}
\partial_{t_1}\frac{p}{c_s^2} +\nabla\cdot \bm u&=0,\\
\partial_{t_1}u_x +\partial_x (u_x^2+p)+\partial_y (u_y u_x)&=\frac{F_x^{(1)}}{\rho},\\
\partial_{t_1}u_y+\partial_x(u_x u_y)+\partial_y(u_y^2+p)&=\frac{F_y^{(1)}}{\rho},
\end{aligned}
\end{equation}
and the elements of Eq.~(\ref{eq:Mg1}) with $i=1,7,8$ can be written as,
\begin{equation}\label{eq:t2momentum_auxilliary}
\begin{aligned}
-s_{1}^g\left( m_{g_1}^{(1)}+\frac{\tilde{G}_1^{(1)}}{2} \right)&=
\partial_{t_1}\left(3u^2-6p\right)-\tilde{G}_1^{(1)}, \\
-s_{7}^g\left( m_{g_7}^{(1)}+\frac{\tilde{G}_7^{(1)}}{2} \right)&=
\partial_{t_1}\left(u_x^2-u_y^2\right)+\partial_x \left(\frac{2}{3} u_x\right)
-\partial_y\left(\frac{2}{3}u_y\right)-\tilde{G}_7^{(1)},  \\
-s_{8}^g\left(m_{g_8}^{(1)}+\frac{\tilde{G}_8^{(1)}}{2}\right)&=\partial_{t_1}\left(u_x u_y\right)+\partial_x\left(\frac{1}{3}u_y\right)
+\partial_y\left(\frac{1}{3} u_x\right)-\tilde{G}_8^{(1)}.\\
\end{aligned}
\end{equation}
Similarly, substituting Eqs.~(\ref{mg0}-\ref{mG1}) into Eq.~(\ref{eq:Mg2}), the elements of  Eq.~(\ref{eq:Mg2}) with $i=0,3,5$ can be written as,
\begin{equation}\label{eq:epsilon2-0-mac}
\partial_{t_2}\frac{p}{c_s^2}=0,
\end{equation}
\begin{equation}\label{eq:epsilon2-3-mac}
\partial_{t_2} u_x+\partial_x \left[\frac{1}{2}s_1^g \xi \left(m_{g_1}^{(1)}+\frac{\tilde{G}_1^{(1)}}{2}\right)
+\frac{3}{2}s_7^g \nu \left(m_{g_7}^{(1)}+\frac{\tilde{G}_7^{(1)}}{2}\right) \right]
+\partial_y \left[3s_8^g \nu \left(m_{g_8}^{(1)}+\frac{\tilde{G}_8^{(1)}}{2}\right) \right]=\frac{F_x^{(2)}}{\rho},
\end{equation}
\begin{equation}\label{eq:epsilon2-5-mac}
\partial_{t_2} u_y+\partial_x \left[3\nu s_8^g \left(m_{g_8}^{(1)}+\frac{\tilde{G}_8^{(1)}}{2}\right)\right]+\partial_y\left[\frac{1}{2} s_1^g \xi\left(m_{g_1}^{(1)}+\frac{\tilde{G}_1^{(1)}}{2}\right)
-\frac{3}{2}s_7^g \nu\left(m_{g_7}^{(1)}+\frac{\tilde{G}_7^{(1)}}{2}\right) \right]=\frac{F_y^{(2)}}{\rho},
\end{equation}
where $\nu=c_s^2(\frac{1}{s_{7,8}^g}-\frac{1}{2})\delta t$ represents the kinetic viscosity and $\xi=c_s^2(\frac{1}{s_1^g}-\frac{1}{2})\delta t$ represents the bulk viscosity.
With the help of Eq.~(\ref{eq:t2momentum_auxilliary}), Eqs.~(\ref{eq:epsilon2-3-mac}) and (\ref{eq:epsilon2-5-mac}) can be reduced to
\begin{equation}\label{eq:simplifyt2}
\begin{aligned}
\partial_{t_2} u_x&=\partial_x \nu (2\partial_x u_x)+\partial_y \nu(\partial_x u_y+\partial_y u_x) +\partial_x (\xi-\nu )\partial_{\alpha}u_{\alpha} +\frac{F_x^{(2)}}{\rho},\\
\partial_{t_2} u_y&=\partial_x\nu(\partial_x u_y+\partial_y u_x)+\partial_y \nu(2\partial_y u_y)+\partial_y (\xi-\nu) \partial_{\alpha}u_{\alpha}
+\frac{F_y^{(2)}}{\rho}.
\end{aligned}
\end{equation}
By assembling Eq.~(\ref{eq:epsilon1-035-mac}) at the $\epsilon$ scale and Eq.~(\ref{eq:simplifyt2}) at the $\epsilon^2$ scale, the recovered macroscopic equations are
\begin{equation}
\partial_t \frac{p}{c_s^2}+\nabla \cdot \bm u=0,\\
\end{equation}
\begin{equation}
\begin{aligned}
\partial_t \bm u+ \nabla \cdot (\bm u\bm u)=
-\nabla p+ \nabla \cdot [\nu(\nabla\bm u+\nabla\bm u^T)]
+\nabla\cdot [(\xi-\nu)(\nabla\cdot\bm u)]+\frac{\bm F}{\rho}.
\end{aligned}
\end{equation}
In the nearly incompressible limit, the time derivative of the pressure is small and the  divergence-free condition is approximately satisfied. Thus Eq~(\ref{eq:distribution_NS}) can exactly recover the hydrodynamic equations.
\section{Selective filters}
\label{selectiveFilter}
Coefficients of the explicit centered selective filters (SF):
\begin{table}[htbp]
\begin{tabular}{ccccc}
\hline
 & SF-5 & SF-7  & SF-9 & SF-11 \\
\hline
$d_{0}$   & $6/16$     &$5/16$        &0.243527493120     & 0.215044884112\\
$d_{1}$   &$-4/16$     &$-15/64$      &-0.204788880640    &-0.187772883589   \\
$d_{2}$   &$1/16$      &$3/32  $      &0.120007591680     &0.123755948787 \\
$d_{3}$   &          &$-1/64$       &-0.045211119360    &-0.059227575576\\
$d_{4}$   &          &            &0.008228661760     &0.018721609157\\
$d_{5}$   &          &            &                   & -0.002999540835\\
\hline
\end{tabular}
\end{table}
\end{appendix}

\section*{References}
\bibliographystyle{unsrt}
\bibliography{bib} 
\end{document}